\documentclass{ws-p8-50x6-00}
\usepackage{graphicx}  
\usepackage{psfig}

\begin{document}

\title{Fueling the AGN}

\author{F. Combes}

\address{Observatoire de Paris, DEMIRM, 61 Av. de l'Observatoire, F-75 014, Paris,
FRANCE\\E-mail: francoise.combes@obspm.fr}


\maketitle

\abstracts{ Active Galactic Nuclei are fueled from material (gas or stars) that
are in general far away from the gravitational influence of the central black
hole, the engine thought to be responsible for their activity. The required material
has a lot of angular momentum, that is a priori quite difficult to evacuate.
The various dynamical mechanisms that may play a role in this game are
reviewed, including $m=2$ perturbations (bars and spirals),   $m=1$ perturbations
(spirals, warps, lopsidedness), and tidal interactions between galaxies and
mergers. In the latest stages of the merger, a binary black hole could be
formed, and its influence on the dynamics and fueling is discussed. 
Starbursts are often associated with AGN, and the nature of their
particular connection, and their role in the nuclear fueling is described.
Evolution of the fueling efficiency with redshift is addressed.
}

\section{Introduction}

 Some galaxies are dubbed ``active'', which means that they
radiate much more energy than can be sustained by a normal
evolution rate, or average  gas consumption rate, given their size
and gas reservoir mass. Their activity could be due to
a starburst, or to a material-accreting compact nucleus. Both
are often related or simultaneous.  The activity has a short duration
with respect to the galaxy life-time, but can be recurrent.
 A key issue is to understand the mechanisms that trigger the
activity, and how a large amount of material can be funneled
to the central regions, to fuel this activity. 

A first possibility is provided by the dense star clusters that
are present near the nucleus. These nearby stars can provide
gas fuel to the nucleus through their mass loss rates, or
even the tidal disruption of stars can liberate their whole
gas mass. We will present and discuss the various physical
phenomena involved, and their respective time-scales in section 2:
only those stars that have low angular momentum orbits are
available to fuel the activity, but these can be replenished
through dynamical diffusion when depleted. 
A bigger problem arises when the nearby stars are not numerous
enough to sustain the activity: mass has then to come from the
whole galaxy in a short time-scale, and this rises the
problem of angular momentum transfer, that must involve
internal or externally-trigerred gravitational instabilities.
Note that the mass fueling can occur in two steps: a first
instability drives the gas towards the center, giving rise
to a nuclear starburst and the formation of dense nuclear
star clusters. Then, the active nucleus can be fueled by
the evolution of the dense star cluster.

The gravitational instabilities of a galaxy disk are 
described in the next sections. Two-fluid instabilities
are considered, and the critical role of gas is emphasized
(section \ref{interact}). 
Because of its dissipative character, the gas can cool down
as soon as instabilities heat it, and maintain non-axisymmetric
features like spiral structure.The corresponding gravity
torques are the tool to tranfer angular momentum. 
 The most wide-spread instabilities are the $m=2$ spirals
and bars: their formation mechanisms, their family of
orbits, their resonances, etc... are essential to better
understand the dynamical gas fueling (section \ref{bars}).
It will be shown how bars can be destroyed, or re-born,
how bars within bars develop, and how material could
be driven towards the nucleus more efficiently through a 
hierarchy of bars.

Also very frequent are the $m=1$ instabilities, lopsidedness
and off-centring. Several possible mechanisms will be
described for these instabilities, that can also favor
material infall to the center (section \ref{m1}). These
include counter-rotation, warps or peculiar gaseous instabilities
in a near-keplerian disk, a situation common in the neighborhood
of central supermassive black holes. 

Finally, external triggers through galaxy interactions and mergers
should also be emphasized. Observational evidence of
internal and external fueling will be discussed and compared.
 The role of gas and bars in minor and major mergers is
investigated through numerical simulations.
 These external mechanisms are favored by hierarchical
scenarios of galaxy formation, where massive galaxies are
the result of the merging of several smaller entities.  
In this frame, it is expected that the central black hole mass
increases in parallel to the mass of the galaxy, as observations
seem to confirm. 

One consequence of galaxy merging is the possibility of
binary black hole formation. The evolution and history of the black
hole population will be investigated through the main dynamical
mechanisms (section \ref{binbh}), and implications on the
fueling will be explored. The Starburst-AGN connection
(competition or collaboration) is then described,
in a cosmological perspective.

\section{ AGN fueling through dense nuclear star clusters}

\subsection{Fueling: why is there a problem?}

To estimate more quantitatively the fueling problem, 
let us consider the typical luminosity and power of an active
nucleus, that can be of the order or higher than 10$^{46}$ erg/s.
If we assume a mass-to-energy conversion efficiency 
$\epsilon \sim$ 10\%  (L = dM/dt c$^2$ $\epsilon$), then the
mass accretion rate dM/dt should be:

dM/dt $\sim$ 1.7 (0.1/$\epsilon$)  (L/10$^{46}$ erg/s)  M$_\odot$/yr

\noindent If the duty cycle of the AGN is of the order of 
10$^8$ yr, then a total mass up to 2 10$^8$ M$_\odot$ should 
be available.
It is a significant fraction of the gas content of a 
typical galaxy, like the Milky Way!
The time-scale to drive such a large 
mass to the center is likely to be larger than 1 Gyr.

For the mass to infall to the center, it must lose its 
angular momentum. Could this be due to viscous torques?
In a geometrically thin accretion disk, one can consider the 
gas subsonic viscosity, where the viscous stress is modelled 
proportional ($\alpha$) to the internal pressure,
with a factor $\alpha < 1$.
This can only gather in 1 Gyr the gas within 
4 $\alpha$ pc typically (e.g. Shlosman et al 1989, Phinney 1994).
 This shows that viscous torques will not couple the large-scale
galaxy to the nucleus, only the very nuclear regions could
play a role through viscous torques.

\subsection{Are stars a possible fuel?}
\label{coeval}

The stars themselves could provide gas to the nucleus, through
their mass loss, if there is a local stellar cluster, dense and compact
enough (core radius R$_c$ of less than a pc, core mass  
M$_{core}$ of the order of 10$^{8}$ M$_\odot$). 
However, the mass loss rate derived from normal stellar evolution gives
only 10$^{-11}$ M$_\odot$/yr/M$_\odot$,
orders of magnitude below the required rate of a few  M$_\odot$ /yr.
 The contribution will be significant, only
if a massive stellar cluster (4 10$^9$ 
M$_\odot$) has just formed through a starburst
(Norman \& Scoville 1988).
A coeval cluster can liberate 10$^9$ M$_\odot$ on 10$^8$ yr,
since mainly massive stars evolve together in the beginning.
 Thus the existence of a starburst in the first place solves
also the problem of the AGN fueling, as in the 
symbiosis model of Williams et al (1999). The angular momentum
problem is now reported to the starburst fueling, and could
be solved only through large-scale dynamical processes. 

There are several processes to fuel gas to the black hole, directly from 
the stars, and these could work for the low-luminosity end of AGNs;
one can invoke:
\begin{itemize}
\item Bloated stars, a phenomenon that makes mass loss more efficient
 (Edwards 1980, Alexander \& Netzer 1994, 1997),
\item Tidal disruption of stars (Hills 1975, Frank \& Rees 1976),
\item Star Collisions (Spitzer \& Saslaw 1966, Colgate 1967, Courvoisier 
et al 1996, Rauch 1999).
\end{itemize}

To quantify all these processes, it is important now to define the
corresponding characteristic radii.

\begin{figure}[ht]
\begin{center}
\rotatebox{-90}{\includegraphics[width=.75\textwidth]{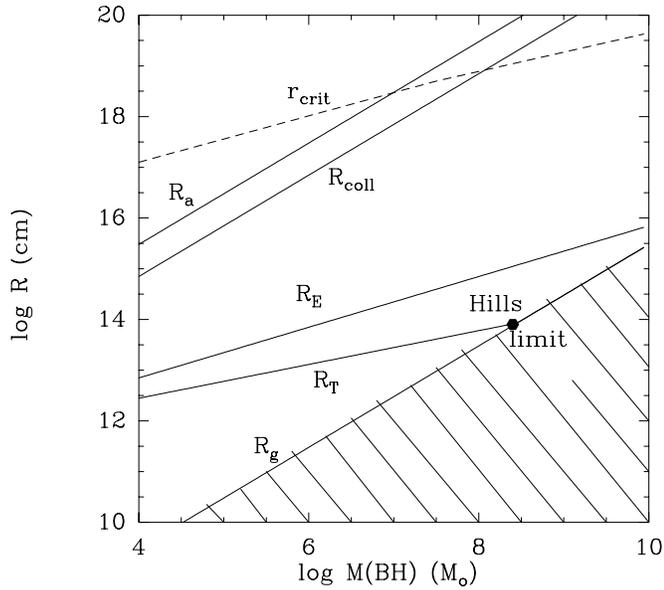}}
\end{center}
\caption{ Characteristic radii, corresponding to the various physical
phenomena, as a function of black hole mass: from top to bottom,
R$_a$, the accretion radius, R$_{coll}$, the collision radius,
R$_E$, the Eddington radius,  R$_T$, the tidal radius, R$_g$, the
gravitational radius (see text for definitions). Loss-cone effects
are important inside the critical radius r$_{crit}$ (see section 2.4). }
\label{char}
\end{figure}

\subsection{Characteristic radii}

First, it is obvious to define the {\bf ``accretion radius''} R$_a$, or the radius
under which the black hole of mass  M$_{\bullet}$
dominates the dynamics, as a function of
the velocity dispersion of the stellar component around it V$_{\infty}$:
$$
R_a = GM_{\bullet}/V_{\infty}^2  \sim 3\times  10^{19} {\rm cm \, M}_8 (200/V_{\infty})^2
$$
where M$_8$ is the black hole mass in units of 10$^8$ M$_\odot$.
The radii are plotted in figure \ref{char} (see also Frank \& Rees 1976,
Luminet 1987).

Then the  {\bf ``collision radius''}  R$_{coll}$, is the
radius under which stellar collisions are 
disruptive, i.e. the freefall velocity around the black hole 
(GM$_\bullet$/r)$^{1/2}$ is comparable to the escape speed v$_*$
of a typical individual star (GM$_*$/r$_*$)$^{1/2}$. For solar mass
stars (escape velocity of the order of 500km/s): 
$$
R_{coll} \sim 7 \times 10^{18} {\rm cm \, M}_8
$$

The  {\bf ``Eddington radius''} is the radius 
under which a star receives more light than its 
Eddington luminosity:
$$
R_E \sim R_* (M/M_*)^{1/2}\sim 7\times 10^{14} {\rm cm M}_8^{1/2}
$$
for solar mass stars. The radiation pressure can then disrupt the envelope,
or at least form bloated stars, more fragile with respect to mass loss.

The {\bf  ``tidal radius''} has a great importance, it is the radius
under which a star is disrupted by the tidal 
forces of the black hole (calculated like a Roche radius):
$$
R_T \sim R_* (M/M_*)^{1/3}\sim 6\times 10^{13} {\rm cm \, M}_8^{1/3} \rho_*^{-1/3}
$$
where  $\rho_*$ is the average density of solar mass stars.
From that, we can estimate the accretion rate due to tidal disruption of stars,
by integrating the mass available (in $\rho_{core} R_T^3$), divided by the
dynamical time, in $\rho_{core}^{-1/2}$:

DM/dt (tide) $\sim$ 0.3 M$_\odot$/yr  (M$_{core}$/10$^8$M$_\odot$)$^{3/2}$ 
(0.01pc/R$_c$)$^{9/2}$ M$_8$

Note that the efficiency of collisions between stars becomes larger than 
the tidal contribution,
for large compactness of the nuclear star clusters, such as their velocity 
dispersion $\sigma_* > v_*$:

DM/dt (coll)  $\sim$ DM/dt (tide) ($\sigma_*/v_*)^4$

Finally, let us recall the horizon radius of the black hole 
(or  {\bf ``gravitational radius''})
under which matter cannot escape:
$$
R_g \sim 2 GM_\bullet/c^2\sim  3 10^{13}  {\rm cm \, M}_8
$$

As the black hole horizon grows faster with  M$_\bullet$ than the 
tidal radius, there is a limit, when M$_8 \sim$ 3, above 
which the star disruption occurs inside the black hole, 
and there is no gaseous release or AGN activity 
(but the black hole grows even more rapidly). This is the
Hills limit (Hills 1975).

\subsection{Black hole growth by star accretion}

Let us compute the time required to reach the critical mass 
M$_c$ where R$_T$ = R$_g$, above which stars are 
swallowed by the black hole without any gas radiation
(M$_c$ = 3 10$^8$ M$_\odot$).
When a tidal breakup of a star (of mass m, radius R) occurs, the 
energy required is taken from the orbital energy of 
the star

E$_b$ = 3/4 G m$^2$/R

\noindent then the gas coming from the disruption will 
have an orbit of typical semi-major axis
$$
r = (2M/3m)  R  /( 1- 2 V^2 R/3GM)
$$
$$
r = 1.5{\rm pc \, M}_8 /( 1 - (<V^2>^{1/2}/535{\rm km/s})^2)
$$
For our own Galaxy, with a 2 10$^6$ M$_\odot$ black hole, this 
means a typical radius of the gas disk of 0.03 pc.

\begin{figure}[ht]
\begin{center}
\rotatebox{-90}{\includegraphics[width=.75\textwidth]{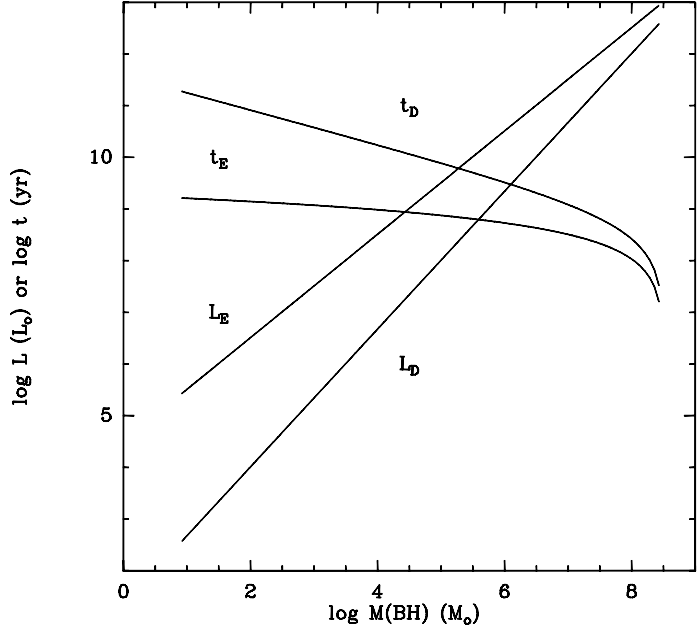}}
\end{center}
\caption{Growth of a supermassive black hole in two simple models:
accretion at Eddington luminosity (time-scale t$_E$ and corresponding 
luminosity L$_E$, as a function of black hole mass M$_{BH}$), and
when accretion is limited by diffusion (t$_D$ and L$_D$)
(from Hills, 1975)  }
\label{hills}
\end{figure}

The black hole cannot swallow the gas too fast, the maximum rate
occurs when it radiates at Eddington luminosity (above which the radiation
pressure prevents the material to fall in). This maximum
luminosity is:
L$_E$ = 3.2 10$^4$ (M/M$_\odot$) L$_\odot$. 
For a mass M$_c$, the maximum is 10$^{13}$ L$_\odot$ (close to the 
peak luminosity of QSOs). Then the corresponding accretion rate, 
assuming an efficiency of $\epsilon$ =10-20\%  is
dM/dt$_E$ = 1.1 10$^{-8}$ (M/M$_\odot$) M$_\odot$/yr.
This implies an exponential growth of the black hole; 
it takes only 1.6 10$^9$ yr to 
grow from a stellar black hole of 10 M$_\odot$ to M$_c$:

t$_E$= 9.3 10$^7$ ln(M$_c$/M) yr

\noindent Note that this very simple scheme would lead to  a 
maximum at z=2.8 of the number of quasars.
This maximum rate, however, is not realistic, since 
the black hole quickly gets short of fuel, as the neighbouring
stars  (in particular at low angular momentum) 
are depleted.  Then it is necessary to consider a 
growth limited by stellar density $\rho_s$:

DM/dt = $\rho_s \sigma$ V

\noindent where $\sigma$ is the accretion cross-section, and
V the typical stellar velocity. The corresponding 
time-scale to grow from M to M$_c$ is

\noindent t$_D$ = 1.7 10$^{15}$ yr ($\rho_s$/M$_\odot$pc$^{-3}$)$^{-1}$ 
M/M$_\odot^{-1/3}$
(1 - M/M$_c^{1/3}$) $<$V$^2>^{1/2}$ (km/s)

\noindent Typically in galaxy nuclei,
 $\rho_s$ = 10$^7$ M$_\odot$/pc$^3$,  $<$V$^2>^{1/2}$  = 225 km/s.
A black hole could grow up to M$_c$ in a Hubble time, and 
the luminosity at the end could be of the order of 
10$^{46}$ erg/s (see figure \ref{hills}).
More detailed considerations
(Frank \& Rees 1976, Lightman \& Shapiro 1977)
introduce the loss-cone effect: the angular momentum can 
diffuse faster than the energy (faster than a stellar relaxation time t$_R$).
Stars with low angular momentum, or very
excentric orbits, will be swallowed first.
Since the low angular momentum stars are replenished faster,
the loss-cone effect  increases the accretion rate by: 
t$_l$ = t$_R$ (1-e$^2$),
with $e$ the excentricity of the orbits.
This is significant inside a critical radius r$_{crit}$, where the loss-cone angle
becomes larger than the diffusion angle $\theta_D \sim (t_{dyn}/t_R)^{1/2}$.
This critical radius is also plotted in figure \ref{char}.

More detailed considerations also can change the above
scenario, for instance when a mass spectrum for the stars is
taken into account.
The critical mass can be then be higher than M$_c$, because 
of large mass stars: giants are less dense and disrupted before
solar-mass stars. This  leads to
higher luminosities for the active nuclei.
Also the presence of the supermassive black hole may form
a cusp of stars in the center. Then the stellar density
is much higher and it is R$_{coll}$ that limits the 
rate of accretion. Gas is produced by the star-star 
collisions, and again higher masses and luminosities 
can be reached.

\subsection{Formation of a cusp of stars around the black hole}

From a numerical resolution of the time-dependent Boltzmann equation,
with the relevant diffusion coefficients, it can be shown that around
a black hole at the center of a globular cluster, the stellar density should
be of a power-law shape, with a slope of 7/4
 (Bahcall \& Wolf 1976). The relevant two-body relaxation time
t$_R$ is dependent on the number of bodies N in the system,
as  t$_R$/t$_c$ = N/logN, where  t$_c$ is the crossing time= r$_c$/V.
For a galactic center, with a volumic density of stars of 
10$^7$M$_\odot$/pc$^3$, this relaxation time is
 3 10$^8$ yr. 

The distribution of stars around a black hole can be described,
according to the distance to the center:
\begin{itemize}
\item first for the stars not bound to the black hole,
at r $>$ R$_a$, their velocity distribution is
Maxwellian, and their density profile has the
isothermal power law in r$^{-2}$. There are also
unbound stars inside R$_a$, but with a density
in  r$^{-1/2}$. This allows to compute the penetration
rate of these unbound stars in the tidal or collision radius,
to estimate the accretion rate. With a core stellar mass of 
M$_{core}$ = 10$^7$- 3 10$^8$ M$_\odot$, a density 10$^7$ pc$^{-3}$,
the accretion rate is, by tidal disruption:  

 dM/dt$_{tide}$ = 1 M$_\odot$/yr M$_8^{4/3}$

\noindent and by stellar collisions: 

dM/dt$_{coll}$ = 0.1 M$_\odot$/yr M$_8^3$

\item the orbits bound to the black hole  r$<$ R$_a$: due to the cusp,
their density is in  r$^{-7/4}$, there is an excess of stars
inside R$_{coll}$, that favors stellar collisions.
\end{itemize}

More refined Monte-Carlo simulations with 
a distribution function f(E,L, t), taking into account the velocity 
anisotropy, the disparition of disrupted stars, etc.. 
show that the stellar cluster cannot fuel the black hole 
indefinitely (Duncan \& Shapiro 1983).
The growth rate of the black hole and its luminosity 
decreases as 1/time.
The loss-cone theory and the simulations are in 
agreement:
the accretion rate due to tidal disruptions  is 
M$_{core}$/t$_R$, typically of 10$^{-2}$ M$_\odot$/yr, with a 
 maximum lower than 1 M$_\odot$/yr; this cannot explain 
the luminosity of QSOs. QSOs might be explained only when stellar 
collisions are included, the corresponding accretion
rate is typically a hundred times higher.

Triaxial deviations from spherical symmetry of 
only 5\%  (due to a bar or binary black hole)
can repopulate the loss-cone, increasing tidal 
disruption to QSOs levels.
However, t$_{coll} <$ t$_R$, and collisions may destroy 
the cusp (Norman \& Silk 1983).

\begin{figure}[ht]
\begin{center}
\rotatebox{0}{\includegraphics[width=.75\textwidth]{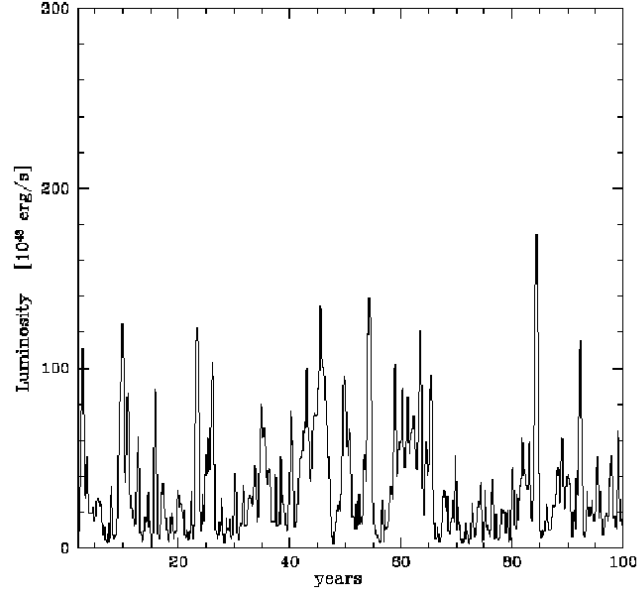}}
\end{center}
\caption{ Light curve produced by the energy released 
in star collisions, with a collision 
rate of $\tau$ = 12 collisions per yr, provided for instance
by a star cluster of mass 10$^{7}$ M$_\odot$,
a core radius of R$_c$ = 0.001 pc, surrounding a black hole of 
mass  10$^{9}$ M$_\odot$
(from Torricelli-Campioni et al 2000).}
\label{courvoisier}
\end{figure}

\subsection{Stellar Collisions}

There is a tight analogy between the tidal disruption of stars 
by the black hole and the disruption of two stars during 
a head-on collision. The 
factor $\beta$ = v$_{coll}$/v$_*$  (v$_*$ = 500km/s for a solar 
mass star) regulates the fate of collisions, and is equivalent
to the factor of penetration of the star inside the tidal
radius R$_T$/r$_{peri}$, where r$_{peri}$ is the pericenter
of the stellar orbit.
\begin{itemize}
\item if $\beta$ = 1, the collision occurs near the
collision radius R$_{coll}$, there is coalescence of the two stars;
 the result may quickly explodes as a supernova.
\item if $\beta >$ 1  a fraction of the mass is lost, and fuels 
the black hole.
\item if $\beta >>$ 1  (well inside R$_{coll}$), a deformation in 
pancakes occurs, very similar to the tidal disruption, when
R$_T$/r$_{peri}>>$ 1.
Gas and debris are bound to the black hole.
\end{itemize}
Stellar collisions help to refill the loss-cone,
although they flatten the stellar cusp (Rauch 1999).
The collisions rate is comparable to the diffusion 
rate, that refill the central core.

It is now generally accepted that
for low density nuclei, stellar evolution and 
tidal disruption is the main mechanism
to bring matter to the black hole, and
for high density nuclei, stellar collisions 
dominate the gas fueling.
The evolution of the stellar density through these processes
is then opposite, and accentuates the differences:
-- for n $<$ 10$^7$/pc$^3$ the core then expands, due to 
heating that results from the settling of a small 
population of stars into orbits tightly bound to 
the black hole;
-- for  n$>$ 10$^7$/pc$^3$, the core shrinks due to the 
removal of kinetic energy by collisions.
To give an order of magnitude, the nuclear density
in our own Galaxy is estimated at
10$^8$ M$_\odot$/pc$^3$ (Eckart et al 1993).

These mechanisms produce differing power-law 
slopes in the resulting stellar density cusp 
surrounding the black hole, -7/4 and -1/2
for low- and high-density nuclei, respectively
(Murphy et al 1991, Rauch 1999).
In simulations however (Rauch 1999), collisions tend 
to produce a flat core, instead of r$^{-1/2}$ law in 
Fokker-PLank studies, which imply isotropy 
(and are unable to treat sparse regions).

Finally, stellar collisions could explain both the high
luminosities (up to 10$^{45}$ ergs/s,
Spitzer \& Saslaw 1966, Colgate 1967), 
and the variability of some QSOs ($\sim$ 10yr scale)
according to Courvoisier et al (1996), see figure \ref{courvoisier}.

But still, even with collisions, the most active quasars 
requiring 100 M$_\odot$/yr remain to be fueled.
May be the gas from stars disruption can be 
stored in a disk,
and suddenly poured in a burst of activity ?
(Shields \& Wheeler 1978).
There might be tight
relations with nuclear starbursts
(Norman \& Scoville 1988, Perry 1992, Williams et al 1999).

\section{Large-scale instabilities}

Since the available fuel near the black hole is not sufficient,
and a large fraction of the gaseous mass of the galaxy disk must be 
involved, it is necessary to rely on violent large-scale instabilities,
that can drive the matter inwards in a  few dynamical times.

\subsection{Gravitational stability}

The conditions of stability of a dissipation-less component
like the stellar one, is relatively well known. High enough
velocity dispersions are required to suppress
axisymmetric instabilities and even spirals, bars or
z-instabilities. The latter provide heating to the medium which 
becomes un-responsive.
The gas is dissipative and has a completely different behaviour;
it is always unstable, and its velocity dispersion is fixed
by regulation and feed-back.

The local stability criterion has been established by Toomre (1964):
stabilisation is obtained through pressure gradients
(velocity dispersion $c$) at small scale, smaller than
the Jeans length:
$\lambda = c t_{ff}  = c^2/G\Sigma$, where $\Sigma$ is the
disk surface density.
At large scale, the rotation stabilises through centrifugal
forces, more precisely the scales larger than 
L$_{crit} = 4 \pi^2 G \Sigma / \kappa^2$
 (where $\kappa$ is the epicyclic frequency).
The Safronov-Toomre criterion is obtained in
equalling $\lambda$ and L$_{crit}$:
$$
Q = {{\kappa c}\over {3.36 G \Sigma}} > 1
$$

For one component, a radial mode ($\omega, k$) in a 
linear analysis obeys the dispersion relation (where $k$ is the wave number):
$$
\omega^2 = \kappa^2 + k^2c^2 -2\pi G\Sigma k =  \kappa^2 +4\pi^2c^2/\lambda^2
 -4\pi^2 G\Sigma /\lambda
$$
which means that
self-gravity reduces the local frequency $\kappa$.
If a stability criterium is easy to derive for one component, the same is
not true for a two-components fluid, since the coupling between the
two makes the ensemble more unstable than each one alone.
The one-component criterion can be applied separately
to the stellar and gas components, where the corresponding values of
$\Sigma$ and $c$ are used, leading to $Q_s$ and $Q_g$. 
Because $c_g << c_s$ however, only a small percentage of mass
in gas can destabilize the whole disk, even when $Q_s > 1$ and $Q_g > 1$.
For two fluids with gravitational coupling, the dispersion relation yields
a criterion of neutral stability $\omega^2 = 0$:
$$
 {{2\pi Gk\Sigma_s}\over{\kappa^2 +k^2c_s^2}} +
 {{2\pi Gk\Sigma_g}\over{\kappa^2 +k^2c_g^2}} = F = 1
$$
which gives directly
an idea of the relative weight of gas and stars in the instabilities
(cf Jog 1996).
At low k (long waves), essentially the stars contribute
to the instability, while at
high k (short waves), the gas dominates.
 The neutral equilibrium requires the simultaneous solution of 
 d$\omega^2$ (k)/dk = 0, which leads to a 
system of  2 polynomial equations 4th and 3rd order in k, and
no analytic criterion can be derived; there are only numerical solutions
in terms of a two-fluid
$Q_{s-g}$ value, which is always lower than the $Q_s$ or $Q_g$
values. Q$_{s-g}$ = 1 defines the neutral stability,
by analogy with the one-fluid treatment, at the fastest growing $\lambda$
F=  2/(1 + Q$_{s-g}^2$).
The stability depends strongly on the gas mass
fraction $\epsilon$ (between 5 and 25\%), and there 
can be sharp transitions from high
to low values of $\lambda$, as the mass fraction increases from
$\epsilon$ = 0.1 to 0.15 for instance.  When $Q_s$ is high and $Q_g$ is low
(a frequent situation, given the dissipation and cooling in the gas),
the main instability is at small-scales.
A gas-rich galaxy (with $\epsilon> 0.25$) is only stable at very low
surface densities (this explains the Malin 1-type galaxies, Impey \& Bothun
1989); the center of early-type galaxies, where $Q_s$ is high, and
$Q_g$ is low (due to low gas velocity dispersion), could be dominated
by the gas wavelength, even at very low gas mass fraction: this can
explain spiral arms in galaxies such as NGC 2841 (Block et al 1996).
 Also interaction of galaxies, by bringing in a high amount of gas, may
change abruptly the fastest growing wavelength from $\lambda_s$ to 
$\lambda_g$, and trigger star-formation.
Even beyond the neutral stability
criterium, when $Q_{s-g} > 1$, a galaxy disk can be unstable with respect
to non-axisymmetric perturbations, such as bars or spirals; in this case
also, the two-fluid coupling increases the instability, i.e. the disk
will form a bar, even if the stars or the gas alone are stable with
respect to such perturbations (Jog 1992).

\subsection{Feedback on the dynamics}

Many photometric and kinematic studies
 have computed the $Q$ values over galactic disks,
and it appears that $Q$ does not follow big variations,
but on the contrary is almost constant over the systems, as
if self-regulating processes were at work.
The stellar dispersion has been studied in our own galaxy and in a few
external galaxies by Lewis \& Freeman (1989) and  Bottema (1993).
The velocity dispersion decreases
exponentially with radius, in parallel to the stellar surface density,
and $Q_s$ is nearly constant as a function of radius, at least
for large galaxies, and vary between 2 and 3 from galaxy to galaxy.

If the stellar disk can be considered as a self-gravitating infinite slab,
locally isothermal ($\sigma_z$ independent of $z$), its density obeys:
$$
\rho = \rho_0 sech^2(z/z_0) = {\rho_0\over{ch^2(z/z_0)}}
$$
\noindent where $z_0$ is the characteristic scale-height of the stellar disk,
given by
$$
z_0 = {\sigma_z^2 \over {2\pi G\Sigma_s(r)}}
$$
\noindent where $\sigma_z$ is the vertical velocity dispersion, and
$\Sigma_s(r) = z_0\rho_0$ is the surface density.
The latter has a general exponential behaviour (e.g. Freeman
1970), with a radial scale-length $h$. The scale-height $z_0$ has been
observed to be independent of radius (van der Kruit \& Searle 1981,
1982, de Grijs et al 1997), and there are only small departures from isothermality in z (van
der Kruit 1988). Then, if the mass to light ratio is constant with radius for
the whole stellar disk, and there is no or little dark matter within the
optical disk, we expect to find a velocity dispersion varying as
$e^{-r/2h}$. This is exactly what is found, within the large uncertainties
(Bottema 1993). Since some galaxies of the sample are edge-on and
others face-on, the comparison requires to know the relation between z
and r projection of the dispersion. In the solar neighbourhood,
$\sigma_z = 0.6\sigma_r$, and this ratio is assumed to be valid in
external galaxies too.

As for the gaseous component, the vertical velocity dispersion
is constant with radius in the outer parts of galaxies, where the rotation
curve is flat (Dickey et al 1990, Combes \& Becquaert 1997): $c_g \approx 6km/s$.
 The behaviour of $\kappa \propto 1/r$ (for a flat rotation curve) is
exactly parallel to the gas surface density $\Sigma_g \propto 1/r$ (e.g.
Bosma 1981), and therefore Q$_g \propto c_g\kappa/\Sigma_g$ is constant with
radius in the outer parts. This strongly suggests a regulation mechanism,
that could maintain the values of $Q$ about constant for both stars and
gas.

The mechanism could be simply gravitational instabilities coupled
with gas dissipation. When the medium is cool enough, so that the $Q$
value is too low, gravitational instabilities quickly provide heating.
 The stellar component cannot cool down and keeps hot, although this is
somewhat moderated by the gravitational action of the gas, and the
young stars born in the cool component.
 The gas is even more sensitive to the heating, but it can
dissipate its disordered motions. The key point is that cooling
encourages dynamical instabilities, and therefore produces heating, which
is how the regulation works (cf Bertin \& Romeo 1988).

It is tempting to relate the gas stability criterium ($Q_g \sim 1$) to
the  threshold for star formation in galaxy disks (cf Kennicutt, 1989).
However, things are certainly less simple, since the gas component
does not form stars as soon as it is unstable and form clouds.
In the outer parts of galaxies, the HI gas that extends much further than the
last radius of star-formation, appears patchy, clumpy, and following
some kind of spiral structure. The outer gas is unstable at any scale.
 This suggests that instabilities are present, at the origin of cloud
formation, and are the regulator of the constant $c_g$ and $Q_g$
(Lin \& Pringle 1987).

 Note that there are several ways to maintain $Q$
constant, and that the stars and gas have chosen two different ways:
the stellar component keeps its scale-height constant, while its velocity
dispersion is exponentially decreasing with radius; the gas keeps its
velocity dispersion constant, while its scale-height increases steadily
with radius (linearly, when the rotation curve is flat). This might be
related to the different radial distribution. The gas does not display
an exponential radial profile, may be due to continued infall or accretion.

\section{Bars}
\label{bars}

Spiral waves, bars, and more generally non-axisymmetric features
in the galaxy mass distribution and potential can produce torques 
and gas radial inflows towards the center. Bars are more long-lived
and usually represent a stronger perturbation, and therefore are
the privileged way to fuel the nucleus dynamically. 
Let us first describe the orbits and resonances in a barred
potential, to better understand the existence and direction
of the torques.

\subsection{Orbits and resonances}

First, let us recall the characteristics of orbits in an axisymmetric
potential $\Phi(r)$ in the plane $z=0$. A circular orbit has an angular
velocity $\Omega^2 = {1\over r}{d\Phi\over dr}$. In linearizing the
potential in the neighborhood of a circular orbit, the motion of any
particle can be expressed in first order by an epicyclic oscillation,
of frequency $\kappa$,
$$
\kappa^2 = {d^2 \Phi\over dr^2} + 3 \Omega^2 = r {d\Omega^2\over dr} +4\Omega^2
$$
The general orbit is therefore the combination of a circle and an
epicycle, or a rosette, since there is no rational relation between
the two periods.

\begin{figure}
{\centering \leavevmode
\epsfxsize=.45\textwidth \rotatebox{-90}{\epsfbox{agn_f4a.ps}}\hfil
\epsfxsize=.45\textwidth \rotatebox{-90}{\epsfbox{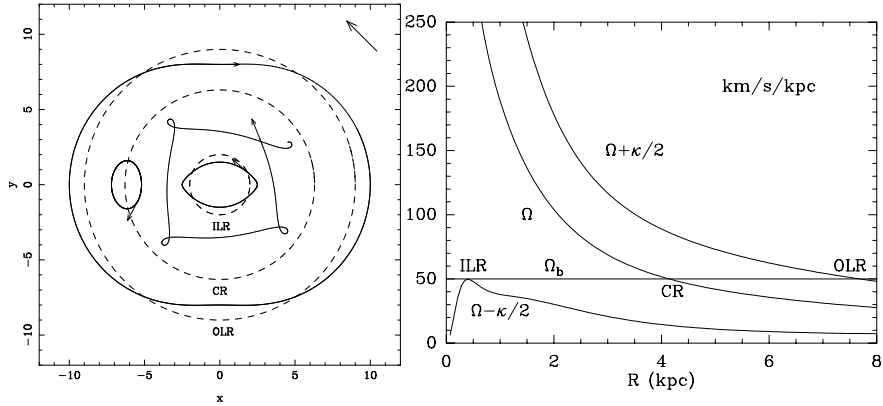}}}
\caption{ {\bf left}: Various types of resonant orbits in a galaxy.
 At the ILR, the orbit is closed and elongated, and
is direct in  the rotating frame. At Corotation (CR), the orbit
makes no turn, only an epicycle; at OLR, the orbit is closed, elongated 
and retrograde in  the rotating frame. In between, the orbits
are rosettes that do not close.
 {\bf right}: Frequencies $\Omega$, $\Omega - \kappa/2$ and $\Omega + \kappa/2$
in a galaxy disk. The bar pattern speed  $\Omega_b$ is indicated, defining the locations
of the Linblad resonances. }
\label{resa}
\end{figure}

The bar creates a bisymmetric gravitational potential, with a predominant
Fourier component $m=2$, which rotates in the galaxy with the pattern speed
$\Omega_b$. There is a region in the plane where the pattern speed
is equal to the frequency of rotation $\Omega$, and where particles
do not make any revolution in the rotating frame. This is the resonance
of corotation (cf fig \ref{resa}).

In the rotating frame, the effective angular velocity of a particle is
$\Omega' = \Omega -\Omega_b$. There exists then regions in the galaxy
where $\Omega' = \kappa/m$, i. e. where the epicyclic orbits close
themselves after $m$ lobes. The corresponding stars are aligned with the
perturbation and closely follow it; they interact with it always with
the same sign, and resonate with it. These zones are the Lindblad
resonances, sketched in Figure~\ref{resa}. According to the relative
values of $\Omega$ and $\kappa$ in a realistic disk galaxy, and because
the bar is a bisymmetric perturbation, the most important resonances
are those for $m=2$. 

Periodic orbits in the bar rotating frame are orbits that close on
themselves after one or more turns. Periodic orbits are the building
blocks which determine the stellar distribution function, since they
define families of trapped orbits around them. Trapped orbits are
non-periodic, but oscillate about one periodic orbit, with a similar
shape. The various families are (Contopoulos \& Grosbol 1989):
\begin{itemize}
\item the $x_1$ family of periodic orbits is the main family supporting
the bar. Orbits are elongated parallel to the bar, within corotation.
They can look like simple ellipses, and with energy increasing, they
can form a cusp, and even two loops at the extremities.
\item  the $x_2$ family exists only between the two inner Lindblad
resonances (ILR), when they exist. They are more round, and elongated
perpendicular to the bar. Even when there exist
two ILRs in the axisymmetric sense, the existence of the $x_2$ family
is not certain. When the bar is strong enough, the $x_2$ orbits
disappear. The bar strength necessary to eliminate the $x_2$ family
depends on the pattern speed $\Omega_b$: the lower this speed, the
stronger the bar must be.
\item  Outside corotation, the 2/1 orbits (which
close after one turn and two epicycles) are run in the retrograde sense
in the rotating frame; they are perpendicular to the bar inside
the outer Lindblad resonance (OLR), and parallel to the bar slightly
outside.
\end{itemize}

\subsection{Gas flow in barred galaxies}

In summary of the previous section, 
the orientation of the periodic orbits rotates by 90$^\circ$ at each
resonance crossing, and they are successively parallel and perpendicular
to the bar.  The gas will first tend to follow
these orbits, but the streamlines of gas cannot cross. Since periodic
orbits do cross, gas clouds can encounter enhanced collisions, such
that their orbits are changed. Instead of experiencing sudden
90$^\circ$ turns, their orbits will smoothly and gradually turn,
following the schematic diagram of kinematic waves, first 
drawn by Kalnajs (1973), and illustrated in fig \ref{fuel1} and \ref{fuel2}.

\begin{figure}
\begin{center}
\rotatebox{-90}{\includegraphics[width=.5\textwidth]{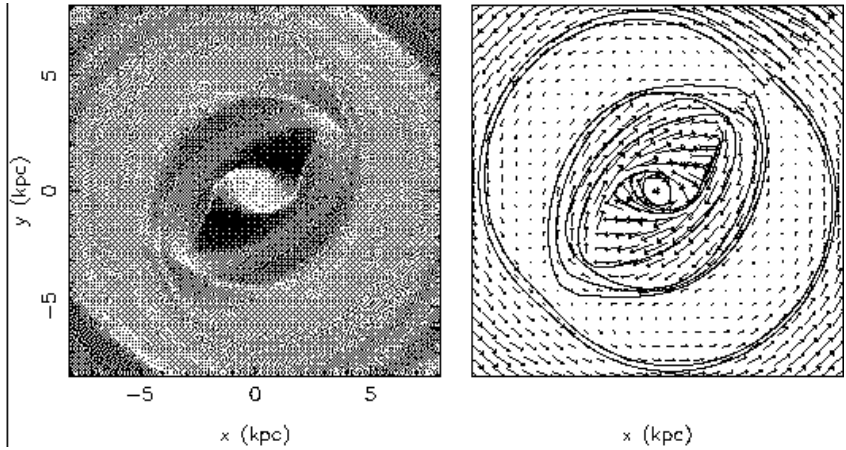}}
\end{center}
\caption{
Response of the gas to a bar potential. The left panel shows the
gas density in grey-scale and the right panel the gas flow lines and
the velocity vectors in the rotating frame of the bar.
The bar potential is at 45$^\circ$ from horizontal
(from Athanassoula, 1992).}
\label{lia}
\end{figure}

This interpretation predicts that the arms will be more wound when there
exist more resonances; there will be a winding over 180$^\circ$ with only
CR and OLR, with the gas aligned with the bar until corotation. When there
exists 2 ILRs, the gas response can be perpendicular to the stellar bar.
When there is barely one ILR, strong shocks can occur on the leading edge
of the bar, corresponding to the offset dust lanes observed in barred
galaxies (cf fig \ref{lia}).

\subsection{Angular momentum transfer}

To minimize its total energy, a galaxy tends to concentrate
its mass towards the center, and to transfer its angular momentum
outwards (Lynden-Bell \& Kalnajs 1972). It is
the role of the spiral structure to transport angular momentum
from the center to the outer parts, and only trailing waves can do it.
This transfer, mediated by non-axisymmetric instabilities,
is the motor of secular evolution of galaxies, and of the formation of
bars and resonant rings.

\begin{figure}
\begin{center}
\rotatebox{-90}{\includegraphics[width=.50\textwidth]{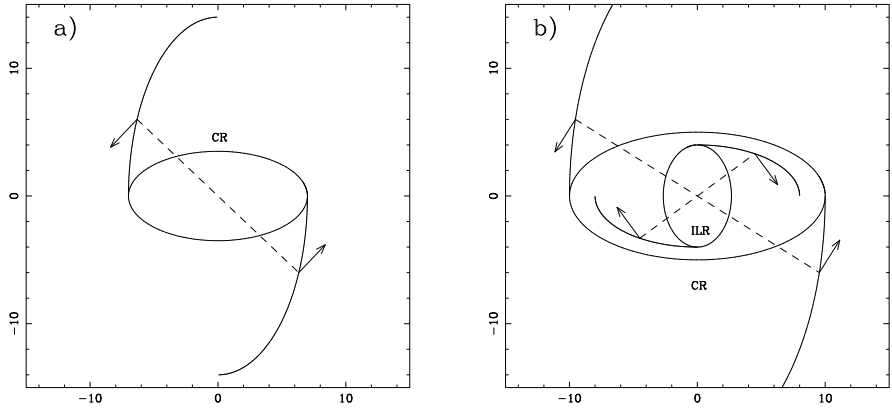}}
\end{center}
\caption{Schematic representation of the gravity torques exerted by
the bar on the gaseous spiral: a) between CR and OLR, the gas acquires
angular momentum and is driven outwards; b) between CR and ILR, the gas
loses angular momentum (cf Combes 1988). }
\label{torq}
\end{figure}

The angular momentum transfer is due to the torques
exerted by the bar on the matter forming spiral arms.
There is a phase shift between the density and the potential
wells, resulting in torques schematized in fig \ref{torq}.
The gas is much more responsive to these torques, since 
they form the spirals in a barred galaxy. 
 The torque changes sign at each resonance, where the
spiral turns by 90$^\circ$.  Between the ILR and corotation, the torque
is negative, while between CR and OLR, the torque is positive.  These
torques tend to depopulate the corotation region, and to  accumulate
gas towards the Lindblad resonances, in the shape of rings. Indeed,
these rings then are aligned with the symmetry axis of the bar, and 
no net torque is acting on them. Numerical simulations of colliding gas
clouds in a barred potential show that rings form in a few dynamical
times, i.e. in a few Gyrs for the outer ring at OLR (Schwarz 1981),
or in $\sim$ 10$^8$ yr for nuclear rings at ILR (Combes \& Gerin 1985).

\begin{figure}
\begin{center}
\psfig{figure=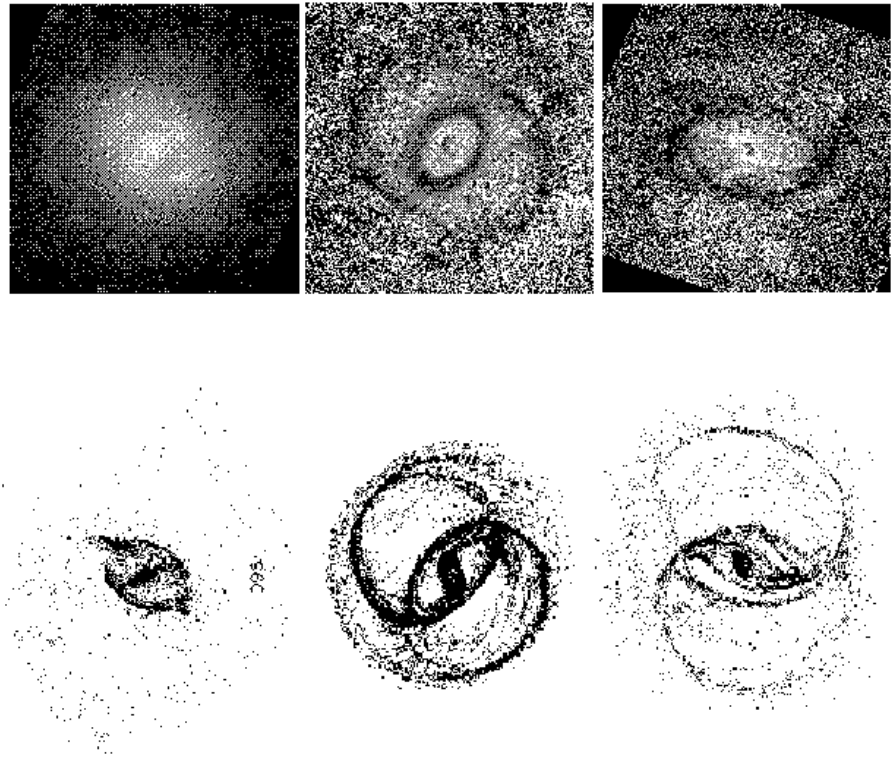,height=12cm,angle=-90,clip=}
\end{center}
\caption{Comparison of observed maps ({\bf top}) and corresponding simulations  ({\bf bottom}) 
for NGC 6300 (left), NGC 3081 (middle) and NGC 1433 (right). For NGC 3081 and NGC 1433,
the top images are B-I colors (blue is dark), and for NGC 6300 it is the H-band image 
(from Buta \& Combes 2000).}
\label{butar}
\end{figure}

This mechanism for ring formation has been confirmed by many studies,
and confronted to observations, where it always encounters large success.
The potential is obtained from photometry of the galaxies in NIR bands,
and derived with a small number of assumptions (usually constant
M/L over the disk). The simulation of the gas flow in the potential 
allow the derivation of the pattern speed (cf Buta \& Combes 1996, 
and fig \ref{butar}). 

\subsection{Fueling nuclear activity}

The main problem to fuel the nucleus is to solve
the transfer of angular momentum problem.
 Torques due to the bar are very efficient, but
gas can be stalled in a nuclear ring at ILR.
Other mechanisms can then be invoked:
viscous torques, or
dynamical friction of giant clouds (GMC) against stars.
The viscosity is in general completely unefficient
over the galactic disk, but the corresponding time-scale 
is decreasing with decreasing radius. Unfortunately,
in the center of galaxies, the rotation is almost
rigid, the shear is considerably reduced, and so are the
viscous torques. 
The time-scale for dynamical friction becomes
competitive below r=100pc from the center (about
10$^7$ (r/100pc)$^2$ yr for a GMC of 10$^7$ M$_\odot$).
 For the intermediate scales, a new mechanism is required.

Note that if there is a supermassive black hole in the
nucleus, it is easier to bring the gas to the center.
Indeed, the presence of a large mass can change the 
behaviour of the precessing rate of orbits $\Omega-\kappa/2$:
instead of increasing with radius  inside ILR (as in fig \ref{resa}),
it will decrease. 

\begin{figure}
\begin{center}
\rotatebox{-90}{\includegraphics[width=.5\textwidth]{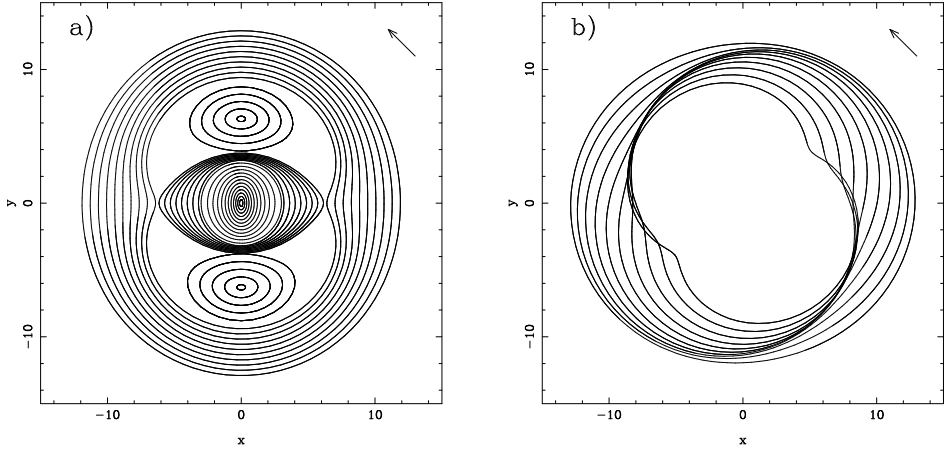}}
\end{center}
\caption{a) Periodic orbits in a barred galaxy ($\cos 2 \theta$ potential,
oriented horizontally). Their orientation rotates by $\pi/2$ at each resonance.
  b) The gas tends to follow these orbits, but is forced to precess more 
rapidly while losing energy and angular momentum, since $\Omega-\kappa/2$
is a decreasing function of radius.}
\label{fuel1}
\end{figure}

\begin{figure}
\begin{center}
\rotatebox{-90}{\includegraphics[width=.5\textwidth]{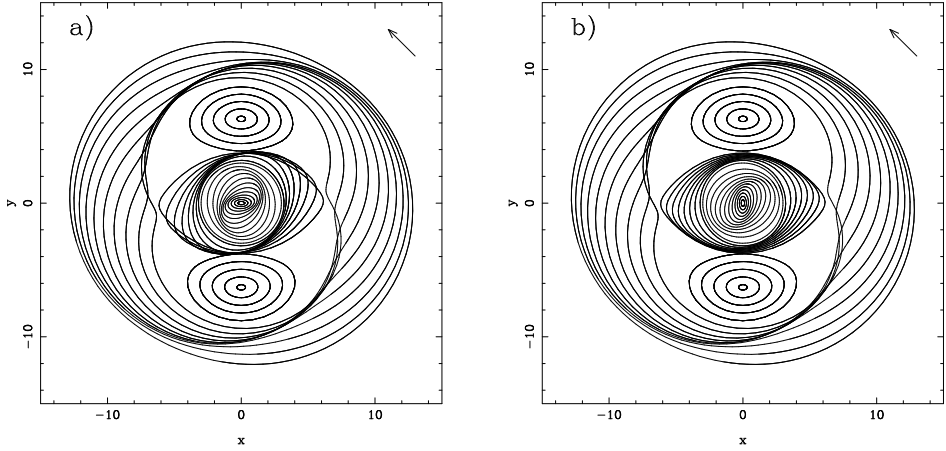}}
\end{center}
\caption{a) Without a central mass concentration, $\Omega-\kappa/2$
is increasing with radius in the center: the gas winds up
in a leading spiral inside the ILR ring; b) with a central mass
concentration, it is the reverse and the gas follows a trailing spiral structure,
inside ILR.}
\label{fuel2}
\end{figure}

Due to cloud
collisions, the gas clouds lose energy, and their galactocentric
distance shrinks. Since it tends to follow the periodic orbits,
the gas streams in elliptical trajectories
at lower and lower radii, with their major axes leading more and more
the periodic orbit, since the precession rate (estimated by $\Omega
-\kappa/2$ in the axisymmetric limit, for orbits near
ILR, and by $\Omega+\kappa/2$ near OLR) increases with decreasing radii
in most of the disk (fig \ref{fuel1}). This regular shift forces the gas into a trailing
spiral structure, from which the sense of the gravity torques can be
easily derived. Inside corotation, the torques are negative, and the
gas is driven inwards towards the inner Lindblad resonance (ILR).
Inside ILR, and from the center, the precessing rate is increasing with
radius, so that the gas pattern due to collisions will be a leading
spiral, instead of a trailing one (see Figure~\ref{fuel2}). The gravity
torques are positive, which also contributes to the accumulation of gas
at the ILR ring. This situation is only inverted in the case of a
central mass concentration (for instance a black hole), for which the precession
rate $\Omega -\kappa/2$ is monotonically increasing towards infinity
with decreasing radii. Only then, the gravity torques will pull the gas
towards the very center, and ``fuel'' the nucleus.

The problem reduces to forming the black hole in the 
first place. We show next that the accumulation of matter towards the center
can produce a decoupling of a second bar inside the primary bar. This
nuclear bar, and possibly other ones nested inside like russian dolls,
can take over the action of gravity torques to drive the gas to the
nucleus, as first proposed by Shlosman et al. (1989).

\subsection{Decoupling of a nuclear disk}

The bar torques drive progressively more mass towards the center.
 This matter, gaseous at the beginning, forms stars, and 
gradually contributes to the formation of the bulge, since
stars are elevated above the disk plane, through vertical
resonances with the bar (e.g. Combes et al. 1990, Raha et al.
1991). When the mass accumulation is large enough,
then the precessing rate $\Omega-\kappa/2$ curve 
is increasing strongly while the radius decreases, and 
this implies the formation of two inner Lindblad resonances.
In between the two ILRs, the periodic orbits are
perpendicular to the bar ($x_2$ orbits), and the bar
loses its main supporters. The weakening of the 
primary bar, and the fact that the frequencies of
the matter are considerably different now between the
inner and outer disk,
forces the decoupling of a nuclear disk, or nuclear bar from
the large-scale bar (primary bar).

\begin{figure}
\begin{center}
\rotatebox{0}{\includegraphics[width=.65\textwidth]{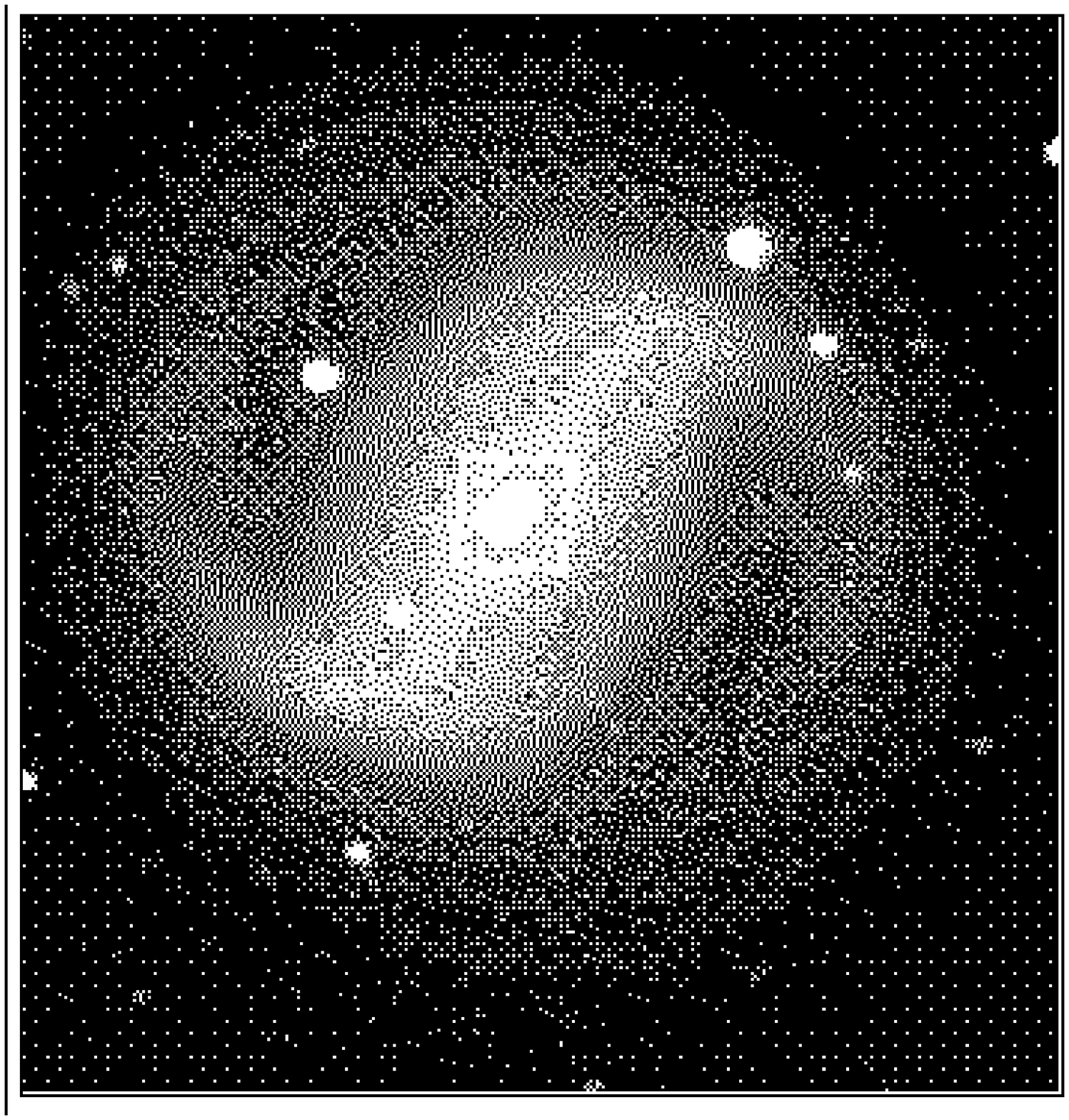}}
\rotatebox{0}{\includegraphics[width=.65\textwidth]{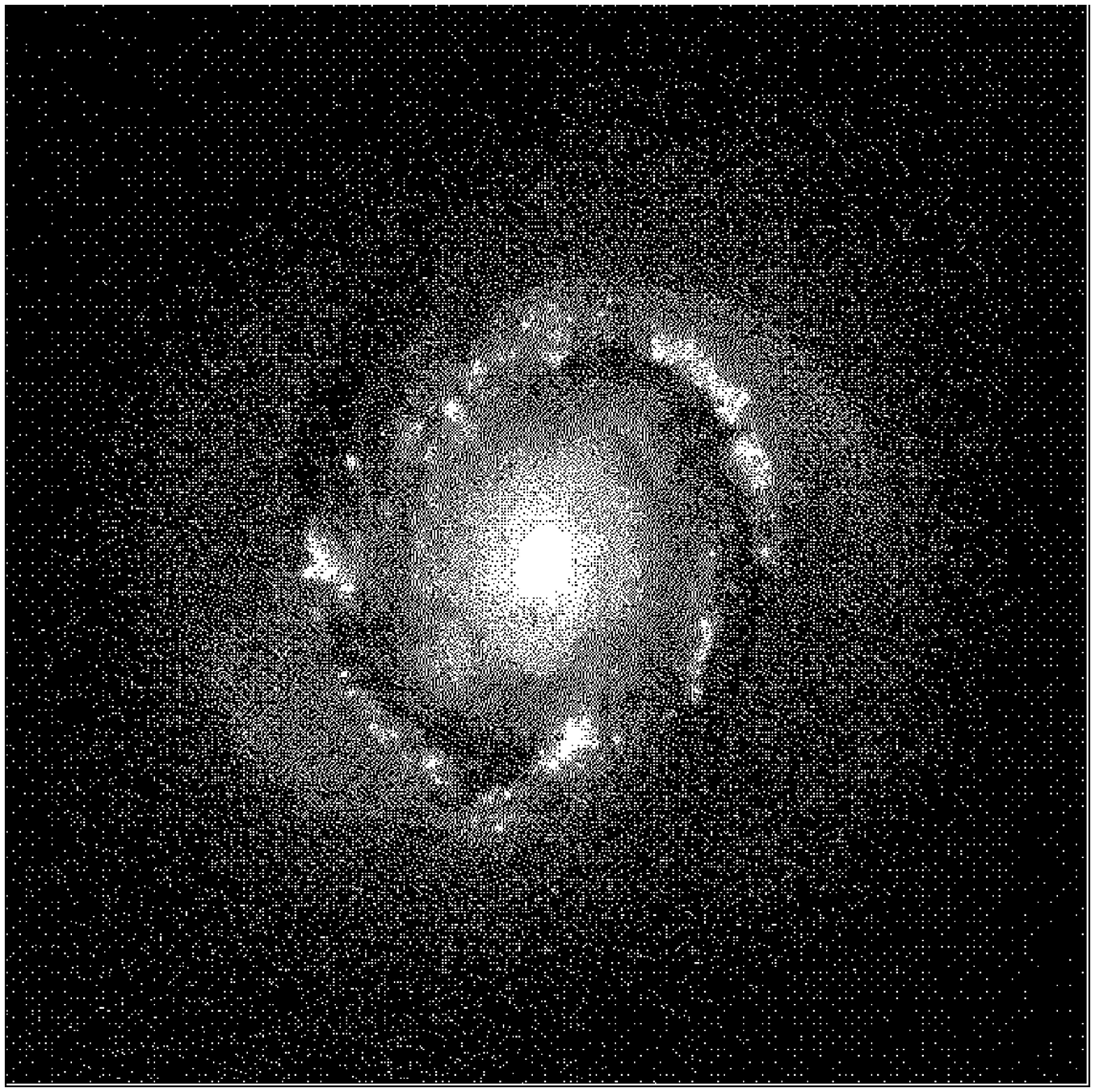}}
\end{center}
\caption{
Photographs of the barred galaxy NGC~4314. ({\bf top})~A general
view of the bar and the spiral arms (photo from McDonald Observatory,
Texas). ({\bf bottom})~A zoom on the central parts (corresp. to the square in the
first picture), with the Hubble Space Telescope.
A nuclear ring can be discerned, within
which a second independent spiral structure has developed
(from Benedict et al. (1996).)}
\label{4314}
\end{figure}

Nuclear disks are frequently observed, at many wavelengths:
optical with the HST (e.g. Barth et al 1995, or fig \ref{4314})
or in CO molecules with millimeter 
interferometers (Ishizuki et al 1990).
A recent survey in the Virgo cluster (Rubin et al 
1997) reveals that about 20\% of the 80 spirals observed 
possess a decoupled nuclear disk. In nearly edge-on
systems, these nuclear disks are conspicuous through
large velocity gradients, like in the Milky-Way (Dame et al 1987)
or NGC 891 (Garcia-Burillo \& Gu\'elin 1995).

\subsection{Bars within bars}

\begin{figure}
{\centering \leavevmode
\epsfxsize=.45\textwidth \epsfbox{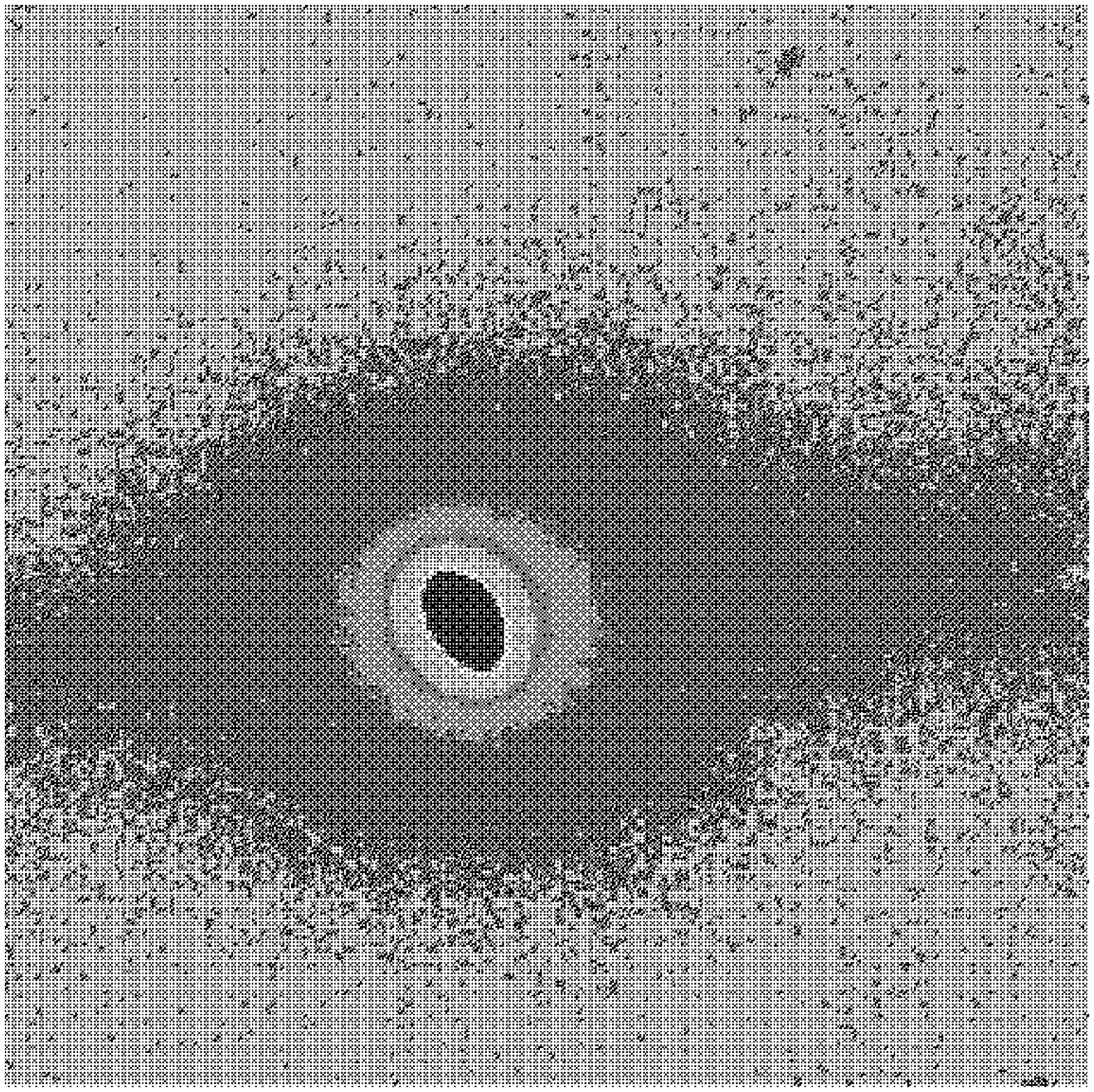} \hfil
\epsfxsize=.45\textwidth \epsfbox{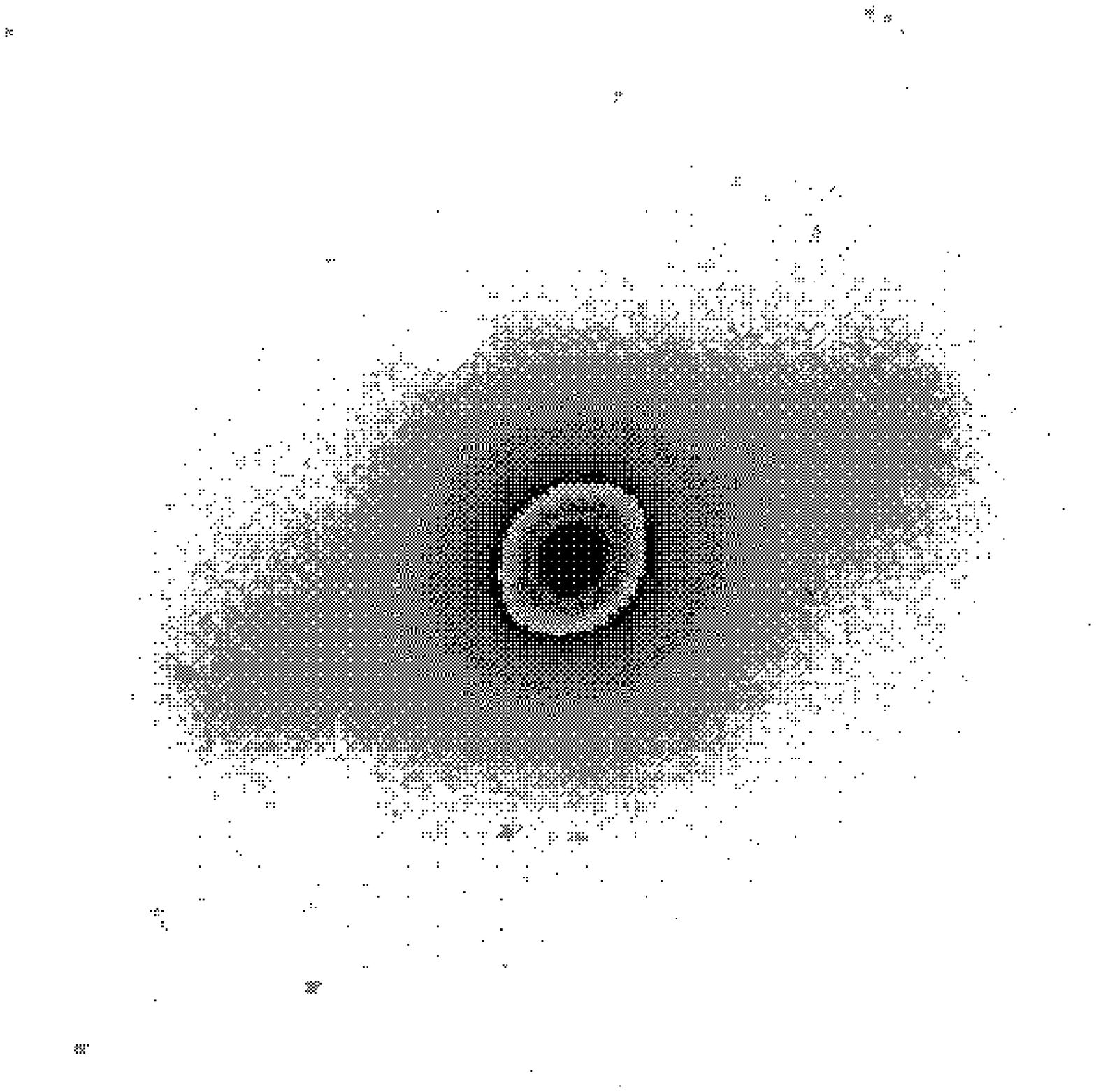}}
\caption{ Examples of nuclear bars seen in the NIR bands:
{\bf left:} NGC 1433 in H; {\bf right:} NGC 2217 in H; the field
of view of both images is 2 x 2 arcmin, from 
Jungwiert et al (1997)}
\label{jung}
\end{figure}

Secondary bars form through decoupling
(Friedli \& Martinet 1993, Combes 1994).
The second bars rotate with a much faster angular velocity,
and are observed with a random angle from the 
primary bar (see fig \ref{jung}).
To avoid chaos, the two bars have a resonance in common.
It is frequent that the ILR of the primary coincides with the
corotation of the secondary. Multiply periodic particle orbits
have been identified in such time-varying potentials
(Maciejewski \& Sparke 1998).
It is possible that the two bars exchange energy with each other,
through non-linear coupling; then $m=4$ and $m=0$ modes are
also expected, and these have been seen in simulations
(Tagger et al. 1987, Masset \& Tagger 1997).
Even then, the life-time of the ensemble is rather
short,  a few rotations. But the nuclear bars could help
to prolonge the action of the primary bar towards the
nucleus (as first proposed by Shlosman et al. 1989).

\subsection{Bar destruction}

The inflow of matter in the center can destroy the bar.
It is sufficient that 5\% of the mass of the disk has sunk inside
the inner Lindblad resonance
(Hasan \& Norman 1990, Pfenniger \& Norman 1990, Hasan et al 1993).
But this depends on the mass distribution, on the size
of the central concentration; a point mass like a black hole is
 more efficient (may be 2\% is sufficient).
 The destruction is due to the mass re-organisation,
that perturbs all the orbital structure: the $x_1$ orbits sustaining
the bar for instance are shifted outwards.
Near the center, the central mass axisymmetrizes
the potential. Then there is a chaotic region,
and outside a regular one again.
 When a central mass concentration exists initially, in
N-body simulations, a bar still forms, but dissolves more
quickly. It is also possible that after a bar has dissolved,
another one forms, after sufficient gas accretion to
generate new gravitational instabilities: the location
of the resonances will not be the same.

\subsection{Nuclear Spirals in Disk Galaxies}

\begin{figure}[ht]
\begin{center}
\rotatebox{0}{\includegraphics[width=.75\textwidth]{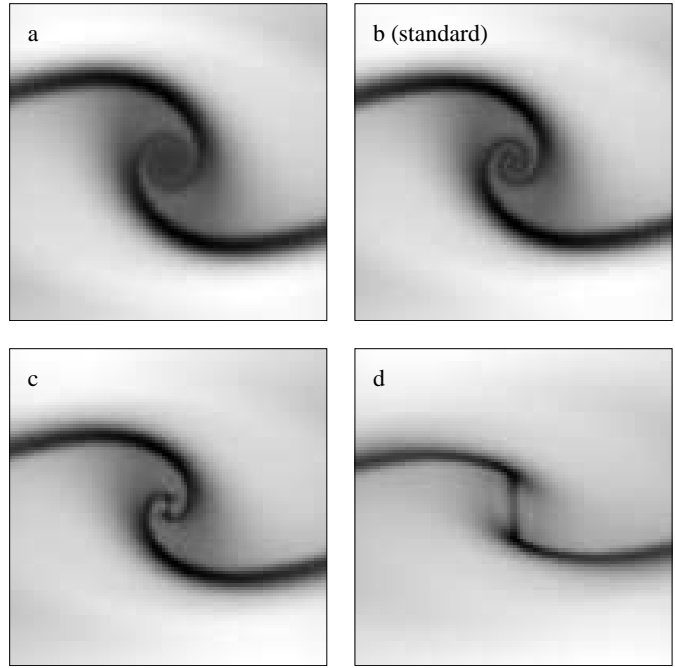}}
\end{center}
\caption{ Gas streaming in a barred potential. The stellar bar is horizontal
and extends more than twice the size of the box. The different morphologies
are due only to the different values of the sound speed: (a) 7 km/s,
(b) 10 km/s, (c) 14 km/s, (d) 20 km/s, from Englmaier \& Shlosman (2000).}
\label{engl}
\end{figure}

 When the mass is concentrated towards the center in 
a galaxy, there can be several possibilities and mechanisms
to fuel the nucleus: either bars within bars, as
multi-mode stellar density waves, as described previously,
or an instable gas-dominated central disk (Heller \& Shlosman 1994).
Another possibility is simply a large-scale stellar bar, and 
a gaseous nuclear spiral by continuity. It has been indeed 
proposed that the observed nuclear spirals are only gaseous
(Elmegreen et al 1998, Regan \& Mulchaey 1999). They could
be acoustic waves, developing without self-gravity.
Gas short waves are able to propagate inside ILR
(and beyond OLR outside).
The dispersion relation is  (with $v_s$ the sound speed):
$$
m^2 (\Omega - \Omega_p)^2 = \kappa^2 +k^2 v_s^2
$$
for m=2
$$
 k^2 v_s^2/4 = (\Omega -\kappa/2-\Omega_p) 
(\Omega +\kappa/2-\Omega_p)
$$
if kR $>>$ 1, the gas waves propagate only when
$(\Omega -\kappa/2 > \Omega_p$,
in between two ILRs, or if there is only one ILR.
Multi-arms (large m) are possible. The value of $Q$ is
large, since the gas is not self-gravitating.
Simulations show the gas response in a stellar bar potential,
and how the morphology depends on the sound speed and
on the shape of the potential (Englmaier \& Shlosman 2000, 
cf fig \ref{engl}).

\subsection{Bars in early- and late-types}

\begin{figure}
\begin{center}
\psfig{figure=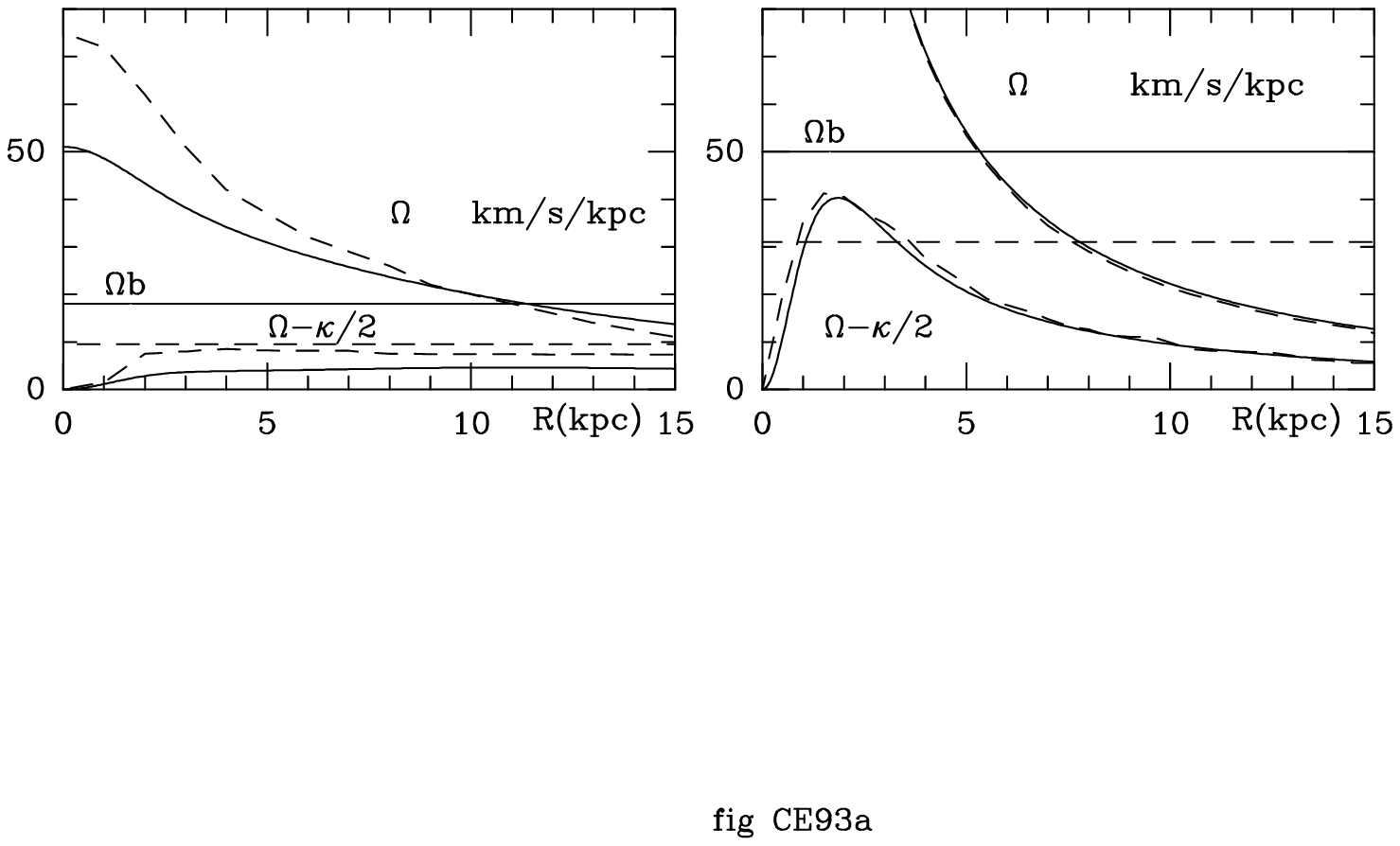,bbllx=60mm,bblly=25mm,bburx=115mm,bbury=180mm,height=12cm,angle=-90,clip=}
\end{center}
\caption{Rotation curves for late and early-type galaxy models,
compared with the bar pattern speed. The initial values are full lines, while
the final values are dashed lines. Left: late-type model; Right: early-type
(from Combes \& Elmegreen 1993).}
\label{ce93a}
\end{figure}

Observations reveal that bars have different properties
along the Hubble sequence
(Elmegreen \& Elmegreen 1985, 1989).
In early-type galaxies, bars have flat profiles,
while in late-types, they have exponential profiles.
Also bars extend farther with respect to the disk exponential 
scale in late-types.
Numerical simulations allow to interpret these features.
Due to their relatively massive bulge, and large mass
concentration, early-types have large precessing frequencies
$\Omega-\kappa/2$ and tend to have large $\Omega_b$
(Combes \& Elmegreen 1993).
There exist then 2 ILRs, and bars end near their corotation.
On the contrary,
late-types have low $\Omega-\kappa/2$, and low $\Omega_b$.
Corotation is then far away in the disk, and lets the disk
scale-length determine the bar length
(see fig \ref{ce93a}).
The profile along the bar is therefore
exponential, very similar to the disk profile.
Early-types bars continuously grow in time,
since angular momentum is transferred in outer regions.
But late-types cannot, since there are not enough
stars to accept angular momentum, 
and they stop to grow before.
 Since the existence of ILRs favors the radial gas inflows,
the fueling appears more efficient in early-type galaxies.

\subsection{Evolution along the Hubble sequence and growth of the black hole}

\begin{figure}[ht]
\begin{center}
\rotatebox{-90}{\includegraphics[width=.75\textwidth]{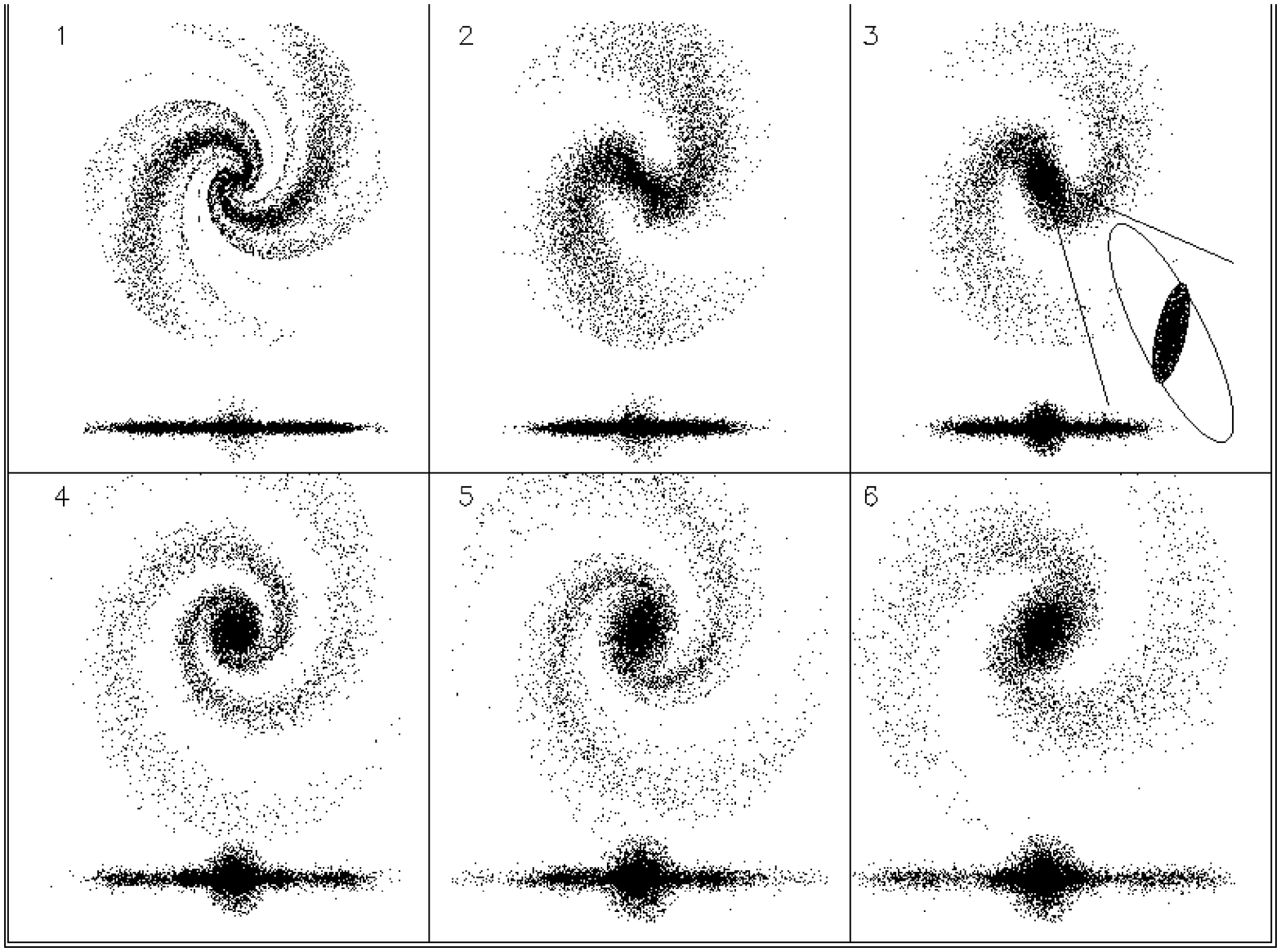}}
\end{center}
\caption{ Cartoon of galaxy evolution along the Hubble sequence.
A galaxy, with a small mass distributed mainly in a disk, without bulge (late-type),
is unstable with respect to spiral and bar formation (steps 1 and 2). 
The bar drives the gas towards the center, and the bulge is building up
(see in each frame the edge-on projection). When there is too much mass
concentrated in the center, the bar is destroyed (step 4), and the gas coming
from the outer parts, enrich the disk, and re-establish a larger disk
to bulge ratio. Later on (steps 5 and 6), another bar will form, when the disk to bulge
ratio is favorable. A secondary bar (cf step 3) may help the primary
one to drive the mass towards the center. At the end, the galaxy may be
classified early-type.}
\label{moriond}
\end{figure}

Non-axisymmetric instabilities like bars force the galaxies to evolve
towards large mass concentrations, large bulge-to-disk ratios, lower
gas content through consumption by star formation. This means
that a galaxy born as a late-type, will progressively evolve towards
early-types, with a somewhat chaotic path, from barred to unbarred, and
sometimes moving backwards, when the disk accretes mass. The gross
lines of this evolution are sketched in fig \ref{moriond}. 

To summarize this bar-driven evolution, and gather the main
features obtained through N-body simulations,
and supported by observations, it is interesting to test a toy model, in a
semi-analytical way, including:

-- star formation, with a combination of
a quiescent rate, proportional to the gas
density, in a time scale of 3 Gyr, and
a bar-driven contribution, with a threshold
(Q$<$1) and a rate equal to (1-Q )/t$_*$,
with t$_*$, the star-formation time-scale (proportional
to the dynamical time-scale for gravitational instabilities).

-- radial flows: when a bar is formed, gravity torques
produce gas inflow, therefore with a threshold
Q$<$1 also,  and rate (1-Q )/t$_{vis}$,
with t$_{vis}$, the ``gravitational viscosity'' time-scale,
 $\sim \frac{1}{\Omega} (\frac{M_{tot}}{M_d})^2$.

-- bulge formation: the inflowing gas (and stars)
are assumed to form the bulge
through star-formation and vertical resonances

-- death of bars: when Q$>$1
  (central concentrations, lack of gas and self-gravitating disk)

-- gas infall: possibility of a continuous small infall  or a
periodically substantial  one (from companions).

-- black hole formation: a fixed fraction b$_{eff}$ of the radial
gas flow is taken to contribute to its formation, i.e.
$ dM_{bh}/dt = b_{eff} M_g (1-Q)/t_{vis}$, with a threshold
Q$<$1.

\begin{figure}[t]
\begin{center}
\psfig{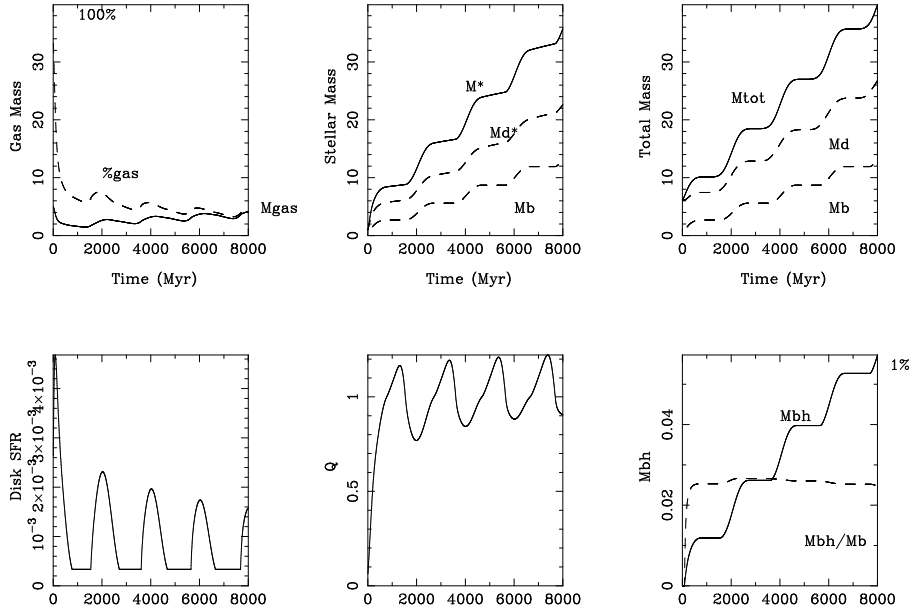}
\end{center}
\caption{ Model of periodic gas accretion in a galaxy with star formation,
birth and death of bars, radial flows and black hole formation taken into
account: {\bf Top left} Full line: gas mass versus time;
dash line: gas mass fraction.  {\bf Top middle} Full line: stellar mass; dash lines: disk
stellar mass at top, and bottom bulge mass.   {\bf Top right} Full line: total mass;
dash lines: total disk mass at top, and bottom bulge mass.
 {\bf Bottom left} Disk star formation rate versus time.  {\bf Bottom middle}
Toomre Q parameter.  {\bf Bottom right}  Full line: mass of the central black hole,
and dash line: mass ratio between the black hole and the bulge.}
\label{visc3}
\end{figure}

Figure \ref{visc3} displays some results of the toy
model (Combes, 2000). The most striking feature is the self-regulation
of the stability parameter $Q$ towards 1. Although the
galaxy initially starts almost completely gaseous, the gas
mass fraction soon stabilises to 10\% of the total. Also
the mass of the central concentration (or black hole)
stabilises to a constant fraction of the bulge mass, as observed
(Magorrian et al. 1998).

\section{m=1 perturbations}
\label{m1}

  Asymmetries, lopsidedness, one-arm spirals are also frequent in galactic
disks, and may play a role in the fueling of the nucleus. Their physical 
processes are varied and different from the more familiar $m=2$
instabilities, and these are first described briefly.

\subsection{Physical nature of instabilities}

The density wave theory has been developped in the WKBJ ($ kr >> 1$ approximation)
and linear regime (e.g. Lin \& Shu 1964, Toomre 1977). The amplification of
the waves occurs at corotation, since the energy and angular momentum
of the perturbation are positive outside and negative inside corotation.
Waves are partially transmitted, and partially reflected at corotation, which
is a zone of evanescence for the waves  if Q $>$ 1.
The wave transmitted will carry energy and angular momentum of 
opposite sign of the incident wave:
for conservation, the reflected wave must have 
increased amplitude. Waves can spontaneously develop if the 
corotation amplifier is coupled to a 
reflection at a resonance or boundary (turning 
point). The feedback cycle may be the WASER (Mark 1974)
or the SWING (Toomre 1977). The turning points are located at the radii when
$\Omega_p$ = ($\Omega \pm \kappa$/m) (1-1/Q$^2$)$^{1/2}$.

\begin{figure}
{\centering \leavevmode
\epsfxsize=.45\textwidth \rotatebox{-90}{\epsfbox{agn_f16a.ps}}\hfil
\epsfxsize=.45\textwidth \rotatebox{-90}{\epsfbox{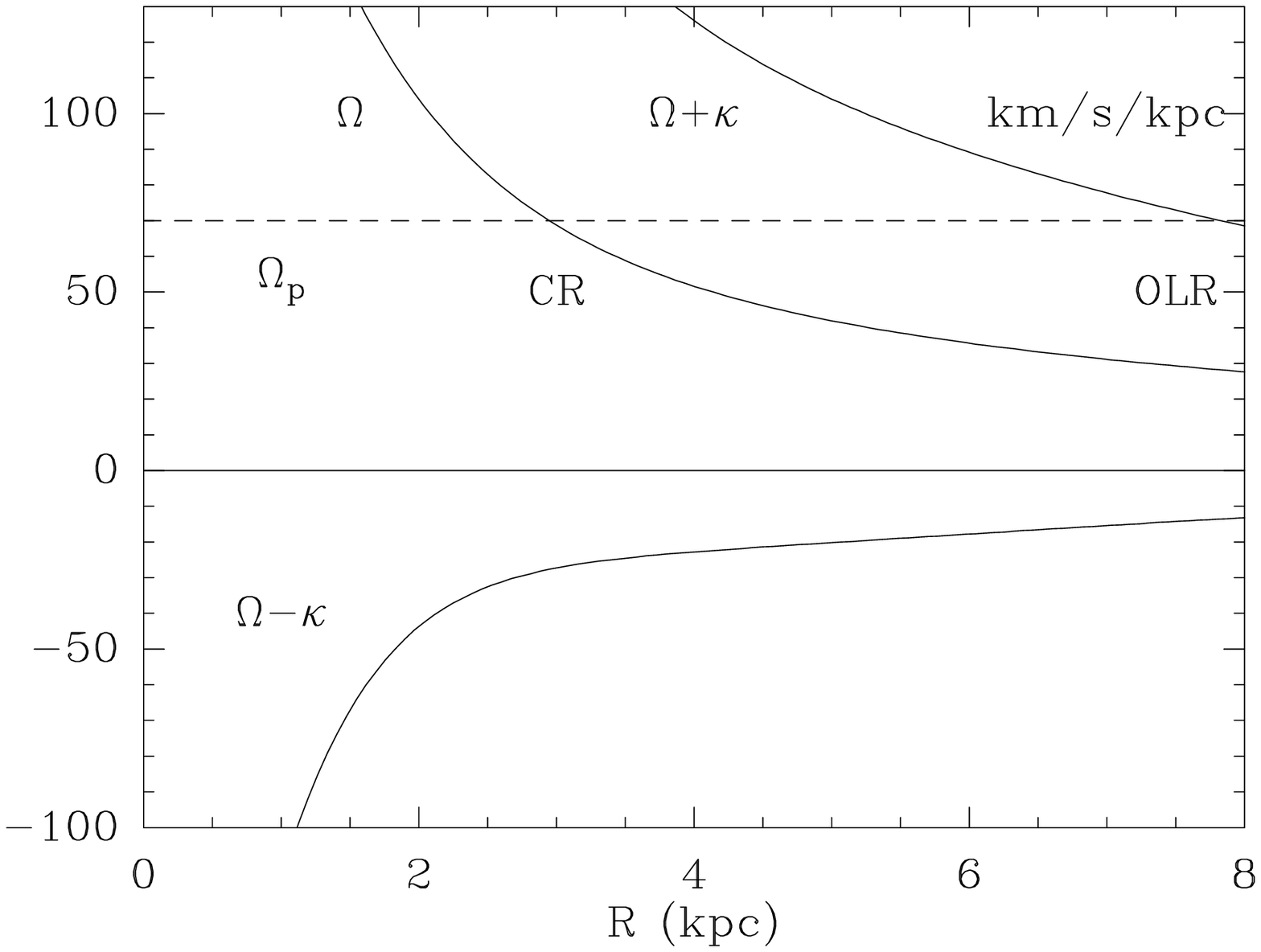}}}
\caption{ {\bf left}: Pattern of lopsided ellipsoidal orbits forming
a one-arm leading spiral.
 {\bf right}: Frequencies $\Omega$, $\Omega - \kappa$ and $\Omega + \kappa$
in a galaxy disk. A possible pattern speed  $\Omega_p$ is indicated, 
allowing CR and OLR resonances}
\label{kspi}
\end{figure}

For $m=1$ perturbations, there cannot exist ILR and OLR at
the same time (see fig \ref{kspi}). 
For lopsided instabilities, developping around a central mass
in a nearly-keplerian disk, there exists another amplifier,
which releases the need of corotation:
the indirect potential, which is due to the off-
centring of the central mass
(Adams et al 1989, Shu et al 1990).
$$
      \Phi (r, \theta, t) = \alpha \omega^2 r \, cos(\omega t -\theta)
$$
This indirect potential creates in permanence a long-range force.
The disk behaves like a resonant cavity
with the off-centring constantly stimulating new 
long trailing waves.
The central mass gains angular momentum, and the disk also
outside corotation: this is not in contradiction, since in 
fact in a frame centered on the central mass, the angular momentum of 
the disk is of opposite sign (with respect to that 
centered on the system center of mass).
While the growth rate of the mode $\gamma$ must be $\sim \Omega$ for the SWING 
mechanism, here $\gamma << \Omega$. 
This mode allows the inner disk to lose angular momentum,
and to inflow on the central mass.

\subsection{Lopsidedness and $m=1$ asymmetries}

The $m=1$ perturbation is present in most galactic disks
(Richter \& Sancisi 1994), often superposed to the $m=2$ ones.
These perturbations can be of different nature and origin,
according to their scale (nuclear or extended disk),
or whether they involve the gaseous or stellar disks.  

Linear analysis, supported by numerical calculations,
show that gaseous disks rotating around a central mass,
are unstable for low value of the central 
mass (Heemskerk et al 1992). The instability disappears
when the central mass equals the disk mass.
 For a gaseous disk, which can develop acoustic waves,
and subject to the indirect term (or off-centering of the center 
of mass) as amplifier, an $m=1$ can form and grow
(Shu et al. 1990, Junqueira \& Combes 1996). 

For nuclear stellar disks, long lasting oscillations of a massive nucleus
have been observed (Miller \& Smith 1992, Taga \& Iye 1998).
For extended disks, off-centered in an extended dark halo,
lopsidedness could survive, if the disk remains in the region
of constant density (or constant  $\Omega$(r)) of the 
halo (Levine \& Sparke 1998).

\begin{figure}
\begin{center}
\rotatebox{-90}{\includegraphics[width=.5\textwidth]{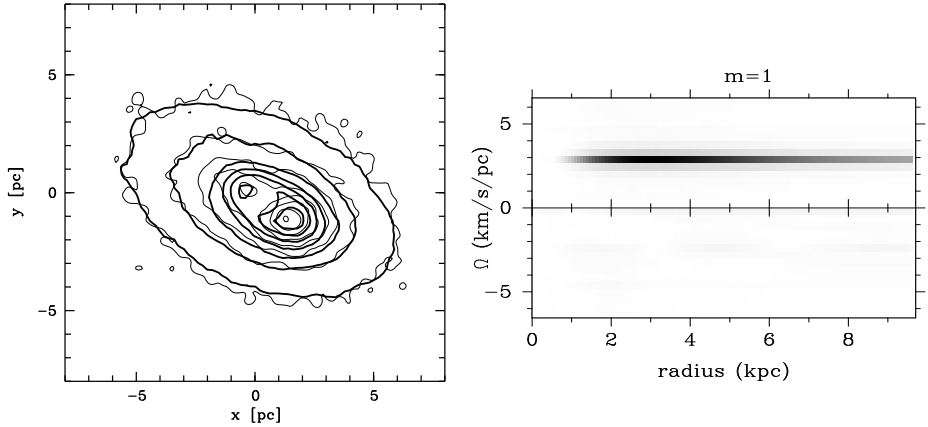}}
\end{center}
\caption{HST map  in the visible of the nuclear disk surrounding 
the black hole of M31: thin lines are
the observed contours, thick lines are those of the m=1 model simulations
({\bf left}).
Pattern speed as a function of radius, for the $m=1$ mode obteined in
the simulation ({\bf right}), from Bacon et al (2000)}
\label{m31}
\end{figure}

In the special case of a stellar nuclear disk around
a massive black-hole, it is possible that self-gravity
is sufficient to compensate for the differential precession
of the nearly keplerian orbits, and that a long-lasting
$m=1$ mode develops, or is maintained long after an
external excitation (Bacon et al. 2000). This could
be the explanation of the double nucleus observed 
for a long time in M31 (Bacon et al. 1994, Kormendy \& 
Bender 1999), and for which an eccentric disk model
has been proposed (Tremaine 1995, Statler 1999).
In this $m=1$ mode, the maximum density is obtained at
the apocenter of the aligned elongated orbits.
The pattern speed, equal to the orbital frequency of the barycentre
of the stellar disk, is slow (3km/s/pc, fig \ref{m31}), 
with respect to the orbital frequency of the stars themselves 
(250km/s/pc in the middle of the disk).
The excitation of the $m=1$ perturbation can then last
more than  3000 rotation periods. This could be a frequent
phenomenon around supermassive black-holes.

The decoupling of the nuclear disks also implies the possibility
of decoupled z-oscillations, and it is frequent to observe these
nuclear regions inclined at a different angle than the main disk.
Jet orientations are not correlated with large-scale disk orientations
in Seyfert Galaxies (Kinney et al. 2000), while they should be
perpendicular to the accretion disk.
As a well-studied example is the Galactic Center,
our nearest supermassive black hole. There must be fueling
in action. There is a large-scale bar, a nuclear bar/spiral, an $m=1$ 
off-centring, a tilted disk (warped?) etc...

\subsection{Counter-rotating components}

The phenomenon of counter-rotating components is
a tracer of galaxy interactions, mass accretion or mergers.
It has been first discovered in ellipticals with kinematically 
decoupled cores (likely to be merger remnants, 
e.g. Barnes \& Hernquist 1992).
It has been observed also in many spirals; either two
components of stars are counter-rotating, or the gas with respect
to the stars, or even two components of gas,
in different regions of the galaxies
(Galletta 1987, Bertola et al 1992). In the
special case of NGC 4550 (Rubin et al. 1992), two
almost identical counter-rotating stellar disks are superposed 
along the line of sight.

These systems pose a number of questions, first from their
formation scenario, but also about their stability, their
life-time, etc.. Do special waves and instabilities 
develop in counter-rotating disks, and 
does this favor central gas accretion?

\subsection{Stability}

First, it appears that the counter-rotation (CR) can bring more stability.
Even a small fraction of CR stars has a stabilising influence with 
respect to bar formation ($m=2$), since the disk has then 
more velocity dispersion (Kalnajs 1977).
But a one-arm instability is triggered, for a comparable 
quantities of CR and normal stars. This comes from the 
two-stream instability in flat disks,
similar to that in CR plasmas (Lovelace et al. 1997). There
develop two $m=1$ modes in the two components,
with energies of opposite signs: the negative-E mode 
can grow by feeding energy in the positive-E mode.

A quasi-stationary one-arm structure forms, and lasts 
for about 1 to 5 periods 
(Comins et al. 1997). The structure is first leading, 
than trailing, and disappears.
The formation of massive CR disks in spirals
has been studied by Thakar \& Ryden (1996, 1998).

\subsection{ Counter-rotation with gas}

The presence of two streams of gas in the same plane 
will be very transient: strong shocks will occur, producing
heating and rapid dissipation. The  
gas is then driven quickly to the center. But the two streams of gas 
could be in inclined planes, or at different radii.
This is the case in polar rings, discovered in
0.5 \% of all galaxies (Whitmore et al. 1990).
After correction for selection effects (non optimal viewing, 
dimming, etc..) 5\% of all S0 would have polar rings.

\begin{figure}
\psfig{figure=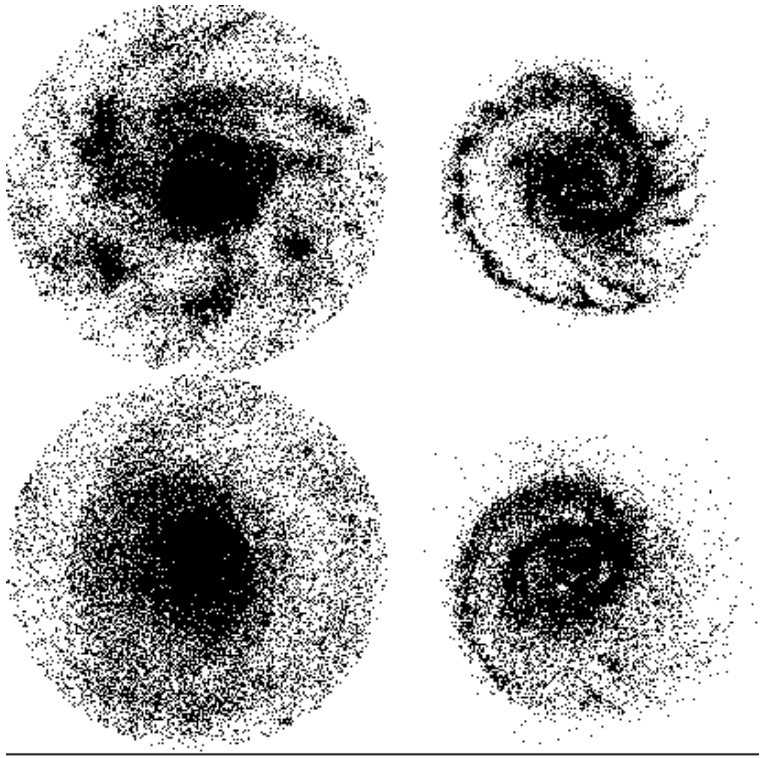,width=10cm,angle=0}
\caption{Particle plots of the stars (left) and the gas (right) for a model of the counter-rotating
galaxy NGC3593,
at successive epochs 200 and 400 Myr. Particles are plotted until a radius of 6.25 kpc.
The majority of stars rotate in the direct sense, while the gas is retrograde (from Garcia-Burillo
et al. 2000)}
\label{cin_ndm}
\end{figure}

\begin{figure}
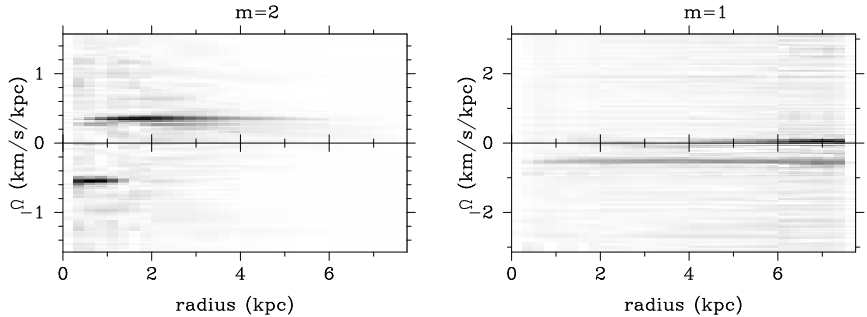

{\centering \leavevmode
\epsfxsize=.35\textwidth \rotatebox{-90}{\epsfbox{agn_f19a.ps}} \hfil
\epsfxsize=.35\textwidth \rotatebox{-90}{\epsfbox{agn_f19b.ps}}}
\caption{Pattern speed as a function of radius, in units of 100km/s/kpc, for
the $m=2$ mode, total density ({\bf left}) and $m=1$ mode, gas density ({\bf right}),
for the NGC3593 model of fig \ref{cin_ndm}}. Note the presence of two counter-rotating
bars.
\label{pow_ndm}
\end{figure}

If there is only one gas stream, the problem is more similar
to the two-stream instabilities of stellar disks mentioned above.
However the gas is cooling, and is not easy to stabilise
against $m=2$ components. Both $m=1$ and $m=2$ may be
present simultaneously in these systems. This is the case of
the galaxy NGC 3593, composed of two CR stellar disks
(Bertola et al 1996): in an extended stellar disk, is 
embedded a CR nuclear disk, possessing co-rotating gas.
 The molecular component associated with this nuclear
disk reveals both a nuclear ring and a one-arm spiral
structure outside of the ring (Garcia-Burillo et al. 2000).
N-body simulations have shown that both structures can
be explained by the superposition of  $m=1$ and $m=2$ 
in the gas component, the ring being formed at the ILR
of the bar (fig. \ref{cin_ndm}). In the $m=2$ pattern, 
two counter-rotating bars develop (fig \ref{pow_ndm}).

\section{Interactions and Mergers}
\label{interact}

Galaxy interactions can be the most efficient way
to produce strong torques, and transfer the angular momentum away.
During the interaction period, strong bars are triggered in the
galaxy disks, and through the same mechanisms as described
before, gas is driven inwards. During the merger, a complete
change of geometry also brings most of the gas to the center.
 The main consequence is the trigger of
spectacular starbursts, in particular in major 
mergers, i.e. merging about equal masses disk 
galaxies. The star forming
activity is concentrated in nuclei, with some 
exceptions (as the antennae or Arp 299 for instance).
 The same gas inflow can also fuel an active nucleus,
either directly, or indirectly through the coeval dense nuclear
clusters formed in the starburst (cf section \ref{coeval}).

In these huge starbursts discovered by IRAS, CO concentrations
suggest that the the cause of the starburst is the
concentration of huge gas masses in the center;
the molecular gas represents a significant 
fraction of the dynamical mass (Scoville et al 1991).
This gas must be brought to the  center in a time-
scale short enough with respect to the feedback 
time-scale of star-formation, i.e. a few 10$^7$yrs
(cf Larson 1987). This requires very 
strong gravity torques.

\subsection{Physical processes}

Numerical simulations have enlightened the dynamical 
processes (Barnes \& Hernquist 1992).
They represent galaxies as 3-components systems,
disks, bulges, and dark haloes. The latter play an
important role, although they do not share most of
the perturbations; they are essential to provide
the dynamical friction that makes the baryonic systems
decay and merge. It is 
the dark matter that takes most of the angular 
momentum away, allowing the luminous mass to 
fall towards the center (Barnes 1988).

Gas dissipation is also a key factor in merger
simulations. The actual equivalent viscosity of the 
ISM is not well known, and it is not possible to reproduce
the full multiphase medium. Two extreme modelisations
are commonly used: the ISM is represented either 
as a continuous fluid, submitted to pressure 
forces, shock waves and artificial viscosity (finite difference
scheme, SPH), or as colliding clouds, with no pressure 
forces and no shocks  (sticky particles). Both can reproduce the main
characteristics of gas flow in galaxies, provided that true 
viscous torques are negligible.
The star-formation rate, global laws and corresponding feed-back
are also not known in most perturbed dynamical situations, and present 
modelisations are rough approximations.
Note that the star formation is itself fundamental
for the dynamics, since it can lock the gas into the non-dissipative
medium and halt for a while the mass transfer towards the center.
 Since stars are observed to be formed inside
giant molecular clouds in our Galaxy, the latter being the result
of agglomerations of smaller entities, one process could be to
relate star formation to cloud-cloud collisions, in the sticky particles
modelisation (Noguchi \& Ishibashi 1986).
Another more widely used is to adopt a
Schmidt law for the SFR, i.e. the rate is proportional to a power $n$ of the
gas volumic density, $n$ being between 1 and 2 (Mihos et al 1992).
 In both cases, it was shown that interacting galaxies were the
site of strong starbursts, that could be explained both by the
orbit crowding in density waves triggered by the tidal interactions,
and by the gas inflow and central concentrations, accumulating the
gas in small and very dense regions. This depends of course on the non-
linearity of the Schmidt law, and SF-efficiency strongly depends on
the power $n$ (see e.g. Mihos et al 1992). Mihos \& Hernquist (1994)
use a hybrid-particles techniques, within SPH, to describe the effects
of gas depletion and formation of a young star population. 

Another peculiar feature of merging galaxies is the ability
of forming giant complexes, that will soon become dense stellar clusters.
This can be understood, given the larger velocity 
dispersion of perturbed systems.
The disks are heated in tidal interactions, since 
the relative orbital energy is transformed into 
internal disordered motions.
This has the consequence to increase the
critical Jeans scale for gravitational instabilities, and to create giant
complexes (e.g. Rand 1993). The global Jeans length is $\lambda
\propto \sigma^2/\Sigma_g$, where $\sigma$ is the velocity dispersion and
$\Sigma_g$ is the gas surface density of the galactic disk. The corresponding
growth time is $\tau_{ff} \propto \sigma/\Sigma_g$, and the instabilities will
occur as soon as Q $\propto \sigma \kappa /\Sigma_g$ becomes lower than 1.
For the same ratio $\sigma/\Sigma_g$, a perturbed system with elevated
$\sigma$ and $\Sigma_g$ will see the condensations of larger complexes
of mass M $\propto \sigma^4/\Sigma_g$, in the same time-scale $\tau$.
These complexes with larger internal dispersions, and larger
gravitational support will be less easy to disrupt through star-formation,
which enhances the star-formation efficiency (Elmegreen et al 1993).
The thermal Jeans length is also larger, due to the hotter gas temperature
induced by the larger number of stars in the clouds, and high mass
stars are favored. This explains the existence of giant and dense star 
clusters observed with HST in merging galaxies
(Schweizer \& Seitzer 1998, Meurer 1995).
These clusters are estimated of masses 10$^7$ M$_\odot$, 
and ages 3 10$^8$ yrs. They could evolve in typical 
globular clusters in about 15 Gyr.

\subsection{Gas flow and starburst/AGN triggering}

\begin{figure}[ht]
\begin{center}
\rotatebox{-90}{\includegraphics[width=.5\textwidth]{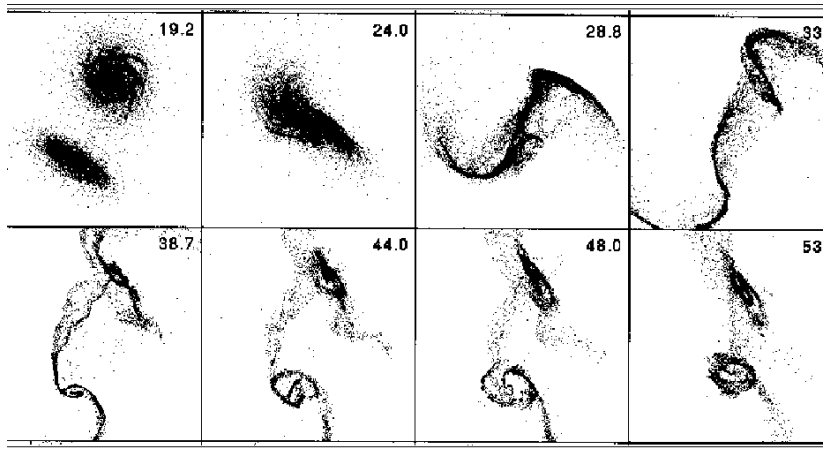}}
\end{center}
\caption{ Simulations of the gas and young stellar components in the
merger of disk/halo models. Time is indicated in units of 12 Myr
(from Mihos \& Hernquist 1996)}
\label{mh96_1}
\end{figure}

\begin{figure}[ht]
\begin{center}
\rotatebox{-90}{\includegraphics[width=.5\textwidth]{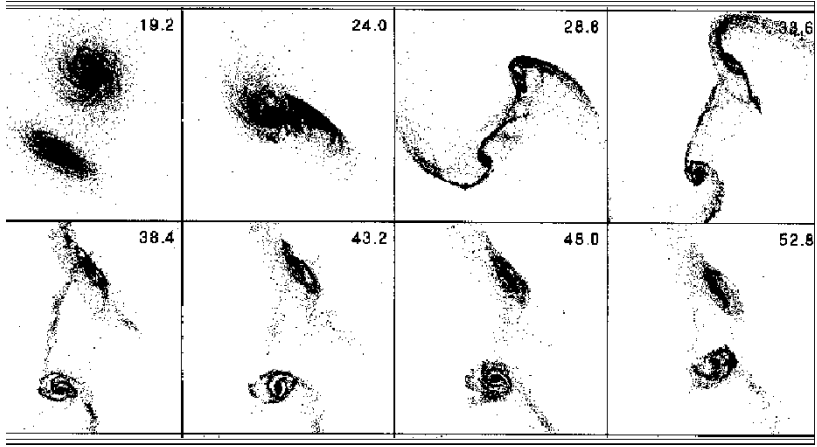}}
\end{center}
\caption{ Same as fig \ref{mh96_1} for the
merger of disk/bulge/halo models }
\label{mh96_2}
\end{figure}

  In major mergers, where the two interacting galaxies are of comparable mass,
strong non-axisymmetric forces are exerted on the gas. But contrary to
what could be expected, the main torques responsible for the gas inflow
are not directly due to the companion. The tidal perturbations destabilise
the primary disk, and the non-axisymmetric structures generated in the
primary disk (bars, spirals) are responsible for the torques.
 The self-gravity of the primary disk, and its consequent
gravitational instabilities, play the fundamental role.
The internal structure therefore takes over from the tidal perturbations
on the outer parts. The gas is provided by the primary disk itself.
Again it is not the gas accreted from the companion which is the
main trigger for activity, until the final merger of course.
  This is why the first parameter determining the characteristics
of the merger event is the initial mass distribution in the two
interacting galaxies (Mihos \& Hernquist 1996). The mass
ratio between the bulge and the disk is a more fundamental
parameter than the geometry of the encounter 
(see fig \ref{mh96_1} and \ref{mh96_2}).

\begin{figure}[ht]
\begin{center}
\rotatebox{-90}{\includegraphics[width=.35\textwidth]{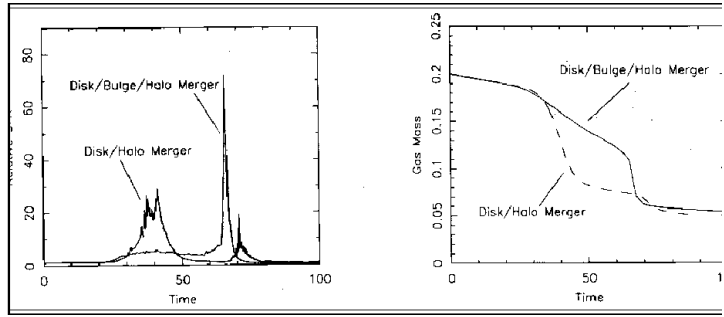}}
\end{center}
\caption{ {\bf left} Evolution of the global star formation rate (relative to isolated disks)
for the two models of fig \ref{mh96_1} and \ref{mh96_2};  {\bf right} Evolution of the total gas mass }
\label{mh96_3}
\end{figure}

 The central bulge stabilises the disk with respect to external
perturbations. If the bulge is sufficiently massive, the apparition
of a strong bar is delayed until the final merging stages, and so
is the gas inflow, and the consequent star-formation activity.
 But the starburst can then be stronger. When the primary disk is
of very late type, without any bulge, the gravitational instability
settles in as soon as the beginning of the interaction, there
is then a continuous activity during the interaction, but
at the end the starburst is less violent, since most
of the gas has been progressively consumed before (see fig \ref{mh96_3}).

The effects of the geometry are more visible on the direct
manifestations of the tidal interaction, i.e. on the tails and debris.
 For the galaxy experiencing a retrograde collision, no extended tidal
tail is formed. There is no resonance effects in the target galaxy disk,
and less violent material perturbations; a transient leading tidal
arm is instead developped. There is however a tidally-induced two-arm
density wave, and the subsequent torques produce gas inflow towards
the center, as for the prograde encounter. During the interaction, it
happens that the retrograde disk has more star formation than the
prograde one, due to the larger gas content. For coplanar merger, the
retrograde disk accretes a significant fraction of the gas from the
prograde disk (Mihos \& Hernquist 1996).

In the simulations, about 75\% of the gas is consumed during the
merger, whatever the internal structure of galaxies, or the
geometry of the encounter (Mihos \& Hernquist 1996). This is
the most uncertain parameter, however.
 What is the fate of the rest of the gas? In general
long tidal tails are entrained in the outer parts,
especially in neutral hydrogen, since this is the
most abundant component in external parts of galaxies.
But most of the material of the tails is still bound
to the system, and will rain down progressively onto
the merger remnant (Hibbard 1995).
Due to phase-wrapping process, it gives rise
to shells and ripples.
The HI fraction in the tails appear to increase as 
the merger enters in more advanced stages.
This might be due to the transformation of HI to 
H$_2$ in the central part.
Also some of the gas is heated to the coronal 
phase (seen in X-rays).

 During minor mergers, where the mass ratio between the two colliding
galaxies is at least 3, the same features can be noticed:
the first relevant parameter is the mass concentration in the
galaxies before the interaction (Hernquist \& Mihos 1995).
The torques responsible for the gas mass inflow are exerted by
the non-axisymmetric (essentially $m=2$)
potential developped in the disk. The torques are much stronger
towards the center; the inflow time is of the order of the
rotation time-scale at each radius.

\section{Observational evidence}

The role of large-scale dynamics and
interactions of galaxies is clear for the starburst 
activity in galaxies
(Kennicutt et al 1987; Sanders et al 1988).
IRAS ultra-luminous galaxies are all mergers
(Sanders \& Mirabel 1996),
and the fraction of interacting galaxies is increasing with 
L$_{IR}$/L$_B$. But the influence of global dynamics is
less clear for AGN activity.
The fraction of Seyfert and quasars is however 
increasing with infrared luminosity,
from 4 to 45\% (Kim et al 1998).

A large fraction of galaxies with radio jets (FRII) 
are interacting (Heckman et al 1986, Baum et al 
1992).
QSOs appear more than usual to interact with 
companions (Hutchings \& Neff 1992; Hutchings 
\& Morris 1995), and 
low-z quasars are mergers on HST images.
 Interactions are therefore efficient, but along years it has
been very difficult to find any correlation between bars
and nuclear activity. Are actually bars responsible for the fueling?

\subsection{ Search for correlations between bars and AGN}

There have been several observational works revealing
a correlation between nuclear activity and bars (Dahari 1984, 
Simkin et al 1980, Moles et al 1995). But the correlation is
weak and depends on the definition of the samples,
their completion and other subtle effects. Near-infrared
images have often revealed bars in galaxies previously 
classified unbarred, certainly due to gas and dust effects.
However, Seyfert galaxies observed in NIR do not statistically
have more bars nor more interactions than a control sample 
(McLeod \& Rieke 1995, Mulchaey \& Regan 1997).

It is however evident from observations 
that bars are efficient to produce radial gas flows:
barred galaxies have more H$_2$ gas concentration
inside their central 500 pc than un-barred galaxies
(cf Sakamoto et al 1999). Also, the radial flows
level out abundance gradients in barred galaxies
(Martin \& Roy 1994).

Peletier et al (1999) and Knapen et al (2000) have recently re-visited 
the question, and took a lot of care with their active and control samples.
Their Seyfert and control samples are different at 2.5 $\sigma$,
in the sense that Seyferts are more barred. They also measure
the bar strength by the observed axial ratio in the images.
In Seyferts, the fraction of strong bars is lower than in 
the control sample (Shlosman et al. 2000).
Although a surprising result a priori, this is not unexpected,
if bars are believed to be destroyed by central mass
concentrations (cf section \ref{bars}).

Regan \& Mulchaey (1999) have studied 12
Seyfert galaxies with HST-NICMOS. Out of the 12, only 3 have 
nuclear bars but a majority show nuclear spirals. However their
criterium for nuclear bars is that there exist leading dust lanes
along this nuclear bar. This is not a required characteristic, since
these secondary bars in general are not expected to have ILRs
themselves. 

A morphological study of the 891 galaxies in the Extended 12 $\mu$m  
Galaxy Sample (E12GS) has confirmed that 
Seyfert galaxies and LINERs have the same percentage of bars as normal spirals,
contrary to HII/Starburst galaxies that have more bars
(Hunt \& Malkan 1999). However, active galaxies show rings significantly
more often than normal galaxies or starbursts. 
The LINERs have more inner rings (by a factor 1.5), while Seyferts have more 
outer rings (by a factor 3-4) than normal galaxies. 
This might be due to the different time-scales for bar and ring formation:
bars form relatively quickly, in a few 10$^8$ yr; they can drive
matter to the central regions and trigger a starburst there, in the
same time-scale. Outer rings form then, also under the gravity torques
of the bar, but in the dynamical times of the outer regions, i.e. 
 a few 10$^9$ yr. Since Seyferts are correlated with them, they would be
associated to delayed consequences of the starburst, or of the bar,
which by this time begins to dissolve.

The percentage of AGN in the E12GS sample is 30\%. As for interactions,
25\% of the Seyferts are "peculiar" (disturbed), while 45\% of the 
HII/Starbursts are. There is also a correlation between AGN and 
morphological types along the Hubble sequence.
Seyferts tend to lie in early-types (Terlevich et al 
1987, Moles et al 1995). This has to be related to the existence 
on inner Linblad resonances in early-types, favoring the fueling
of the nucleus (e.g.  section \ref{bars}, Combes \& Elmegreen 1993).

In summary, there might be some evidence of the role of
large-scale dynamics on AGN fueling, but it is in general weak,
except for the most powerful AGN. 
A good correlation between bars/interactions 
and AGN is not expected, from several arguments:

-- there must be already a massive black hole in the nucleus,
and this might be the case only for massive-bulge galaxies (not all
barred galaxies)

-- again a large central mass concentration (bulge) is necessary to 
produce an ILR and drive the gas inwards  (early-types)

-- other parameters, like geometrical parameters, control 
the fueling efficiency of interaction 

-- time-scales are not fitted: the AGN fueling is 
postponed after the interaction/bar episode

-- there are other mechanisms to fuel AGN such 
that a dense nuclear cluster

\subsection{Galaxy interactions and nuclear activity}

Huge starbursts require galaxy interactions and 
mergers (e.g. Sanders \& Mirabel 1996).
For AGN of low luminosity, external triggering appears less 
necessary, since only 0.01 M$_\odot$/yr is required for 
a Seyfert like NGC 1068 for example, during 10$^8$ yr.
However, once interactions drive gas to the 
nucleus, some activity must be revived.
Time-scales may be the reason why the actions 
are not simultaneous.
Large-scale gas has to be driven at very small 
scales in the center, and the whole process requires
several intermediate steps.

Even for the good starburst/interactions 
correlations, there are exceptions for the low-
luminosity samples.
Interacting galaxies selected optically (not IRAS galaxies) 
are often not enhanced in star formation (Bushouse 
1986, Lawrence et al 1989).
Only the obviously merging galaxies, like the 
Toomre (1977) sample, are truly a starbursting 
class; it is difficult to reveal a progression along a 
possible evolutionary sequence (Heckman 1990).
There are too many determining parameters: 
geometry, distance, gas content, etc... 

A complication comes from the time-scales involved:
the starburst phase is short, of the order of
a few 10$^8$yr, similar only to the end of the merging 
phase. It is more the presence of morphological 
distortions than the presence of nearby 
companions that is correlated with activity.
More than 50\% of ULIRGS possess multiple nuclei
(Carico et al 1990, Graham et al 1990).

Are Seyfert galaxies preferentially interacting? 
According to Dahari (1984), 15\%  have close companions, 
compared to 3\%  in the control sample.
But the Seyferts with or without companions have 
the same H$\alpha$ or radio power (Dahari 1985),
although they may be more infrared bright with companions
(Dahari \& DeRobertis 1988, McKenty 1989).
According to Keel et al (1985),  there are 5\% of Seyferts 
in control sample, and 25\%  in the close pairs of Arp Atlas.
But it is possible that the Arp Atlas galaxies suffer from
selection effects. Bushouse (1986) on the contrary finds a
deficiency of Seyferts in interacting galaxies.

In summary, if the environment influence is evident for
powerful QSOs, Radio-galaxies and BlLacs, it is not so
significant for Seyferts (deRobertis et al 1996).

Surprising observations also involve LSB 
(Low Surface Brightness Galaxies) that are about 6 times
less bright than HSB.  Their low evolution state is generally
attributed to their isolated environment. A large fraction
of active nuclei have been reported in their category: 
65\%  in about 50 LSB instead of 1\% expected for these low luminosities
(Knezek \& Schombert 1993, Sprayberry et al 1995).
Are there unknown selection effects?

\subsection{Radio-galaxies}

Interactions are evident in the more powerful objects only.
The FR-I objects have low-power, radio jets declining with distance,
and are in general elliptical galaxies. Fewer than 10\% show tails, and 
tidal interactions (Smith \& Heckman 1989).
The rarer FR-II objects, with high-power, are classical doubles (Cygnus A, 
Perseus A). They show a high percentage of interactions, 
from 32\%  (Yates et al 1989) to 100\% 
(Hutchings 1987); most of them ($>$ 50\%) have at least 
2 companions, blue colors, star-forming regions
(Heckman 1990, Gubanov 1991).
Against the central radio-source, HI is seen in absorption,
revealing that gas is most of the time infalling (absorption 
lines are blue-shifted, van Gorkom et al 1989).

Radio-loud QSO have 4-5 times more 
neighbors (Yee \& Green 1984, Smith \& Heckman 1990),
while radio-quiet have 2 times more neighbors only with respect
to a control sample. The  morphology of QSO hosts
is disturbed for 35-55\% of them (Hutchings et al 1984, Smith et al 1986),
while radio-loud are perturbed at 70-80\% (Smith et al 1986, 
Hutching 1987); this is confirmed with HST (Disney et al 1995).

The radio power of radio-galaxies has been related to the
spin-down of a rotating supermassive black hole 
(Begelman 1986). The spin of the black hole may be acquired
in galaxy interactions and mergers: either the black hole is spin up
through gas accretion from an external disk, or through the merging
of two black holes. In this respect, it is significant that radio-galaxies
are generally giant ellipticals, since these are expected to be formed
essentially by mergers.

\subsection{Conclusion}

Interactions are an essential mechanism in the 
fueling of QSOs, high-power radio-galaxies and ULIRgs.
But they are not dominant in Seyferts and low-
power radio-galaxies.

Observations reveal a close complicity between starbursts and AGN.
AGN may be fueled by mass loss during PMS 
stellar evolution, of a coeval stellar cluster of 4 
10$^9$ M$_\odot$, within 10pc (Norman \& Scoville 1988).
Within 10$^8$yr, mass loss accumulates 1.5 10$^9$M$_\odot$.
Radiation (UV, X) from the inner accretion disk 
ionizes stellar envelopes, and form the BLR
(V $\sim$ 3000km/s in the black hole potential well).
The young stellar cluster requires a huge 
starburst, like those due to mergers.
Shocks are then inevitable, model-independent
supernovae can help transfering angular momentum.
Starbusrts and AGN are symbiotic systems
(Collin et al 1988, Perry 1992, Williams et al 1999, 
Collin \& Zahn 1999).

\section{Black hole evolution}
\label{binbh}

\begin{figure}
{\centering \leavevmode
\epsfxsize=.40\textwidth \rotatebox{-90}{\epsfbox{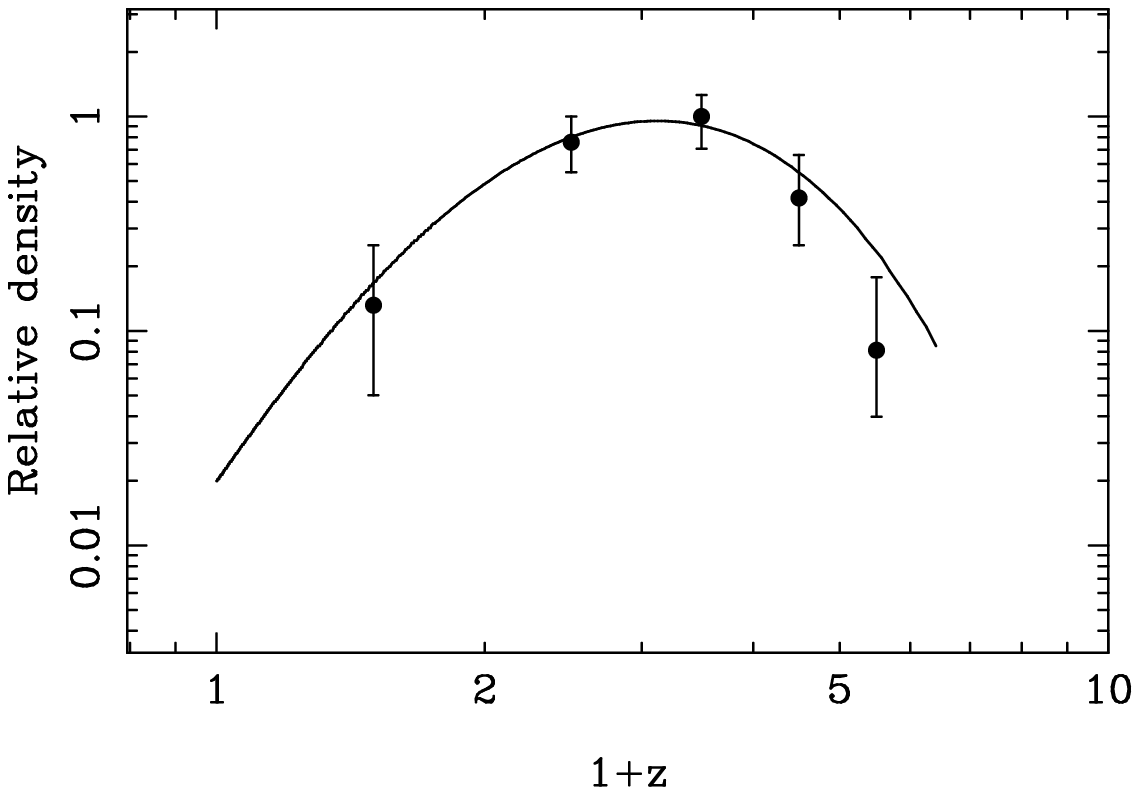}}\hfil
\epsfxsize=.40\textwidth \rotatebox{-90}{\epsfbox{agn_f23b.ps}}}
\caption{ {\bf left} Space density as a function 
of redshift, normalised to $z=2-3$, for the Parkes flat-spectrum
radio-loud quasars with P$_{11} >$ 7.2 10$^{26}$ W Hz$^{-1}$ sr$^{-1}$.
The optically-selected quasars follow the same curve (from Shaver
et al. 1996).
{\bf right} Cosmic history of star formation, for comparison. 
The various data points, coming from different
surveys, give the universal metal ejection rate,
or the star formation rate SFR (left-scale),
as a function of redshift $z$. }
\label{shaver}
\end{figure}

How are formed and grow the massive black holes in the center 
of galaxies? Can we explain the observed peak in quasar activity
at $z=2$ and the decline since then (see fig \ref{shaver})? 
Why is this curve so parallel to the star formation history?
 If galaxies are themselves formed through interactions
and mergers, what is the fate of the black holes in the 
merger, and are binary black holes observable ?
 
\subsection{Black hole growth and activity life-time}

It is widely believed that AGN derive their power 
from supermassive black holes, but their formation
history, their demography, their activity time-scales
are still debated. The two extreme hypotheses have been
explored: either only a small percentage of
galaxies become quasars, and they are continuously fueled,
and active over Gyrs, or a massive black hole exists in
almost every galaxy, but they have active periods of only
a few 10$^7$ yr.  In the first hypothesis, there should exist
blak holes with masses 100 times higher than the maximum
observed today, and accretion rates much lower than 
the Eddington rate, and this is not supported 
(e.g. Cavaliere \& Padovani 1989).
Models with a duty cycle of 4 10$^7$ yr are favored,
and many galaxies today should host a starving 
black hole (Haehnelt \& Rees 1993).

The density of black holes derived today from the kinematics
of galaxy centers at high spatial resolution (Magorrian et al.
1998, Ferrarese \& Merritt 2000) strongly constrains the
 duty cycle time-scale. Also the cosmic background radiation
detected at many wavelengths constrains the formation
history. The volumic density of massive black holes today
$\rho_{bh}$
can be expressed as a function of the density of the bulges
$\Omega_{bul}$ (normalized to the critical density to
close the universe), and of the observed ratio between 
black-hole and bulge mass $M_{bh}/M_{bul}$:
$$
\rho_{bh} = 1.1 \, 10^6 { {M_{bh}/M_{bul}}\over {0.002}}
{ {\Omega_{bul}}\over{0.002h^{-1}} } M_\odot \, Mpc^{-3}
$$  
with $h$ the usual ratio of the Hubble constant to
100 km/s/Mpc (Haehnelt et al. 1998).
 It is interesting to note that this ratio largely exceeds
(by a factor 10) the mass density in black holes needed to 
explain the blue light purely by accretion on AGNs:
$$
\rho_{b} = 1.4 \, 10^5 { {f_b \epsilon }\over {0.01}}
 M_\odot \, Mpc^{-3}
$$  
where $f_b$ is the fraction of light emitted in blue, and 
$\epsilon$ the mass-to-energy conversion efficiency (section 2). 
This might be interpreted as a strong extinction of the blue
light from AGNs, since part of the energy then can be seen 
in the infrared light. From the detected cosmic infrared
radiation, and assuming that about 30\% is coming from AGN
(e.g. Genzel et al 1998), the predicted density is:
$$
\rho_{ir} = 7.5 \, 10^5 { {f_{ir} \epsilon }\over {0.1}}
 M_\odot \, Mpc^{-3}
$$  
Part of the radiation is also observed through hard X-rays, and 
from the background hard-Xray radiation, the corresponding
black hole density would be:
$$
\rho_{x} = 3.8 \, 10^5 { {f_{x} \epsilon }\over {0.01}}
 M_\odot \, Mpc^{-3}
$$  
Note however that the sources contributing to the submillimetric
and hard-Xray backgrounds are not the same, as discovered
recently with Chandra (Severgnini et al. 2000). This could mean
that even the hard-Xrays are extincted and most of the AGN 
radiation comes essentially in the infrared.

\begin{figure}[ht]
{\centering \leavevmode
\epsfxsize=.45\textwidth \rotatebox{-90}{\epsfbox{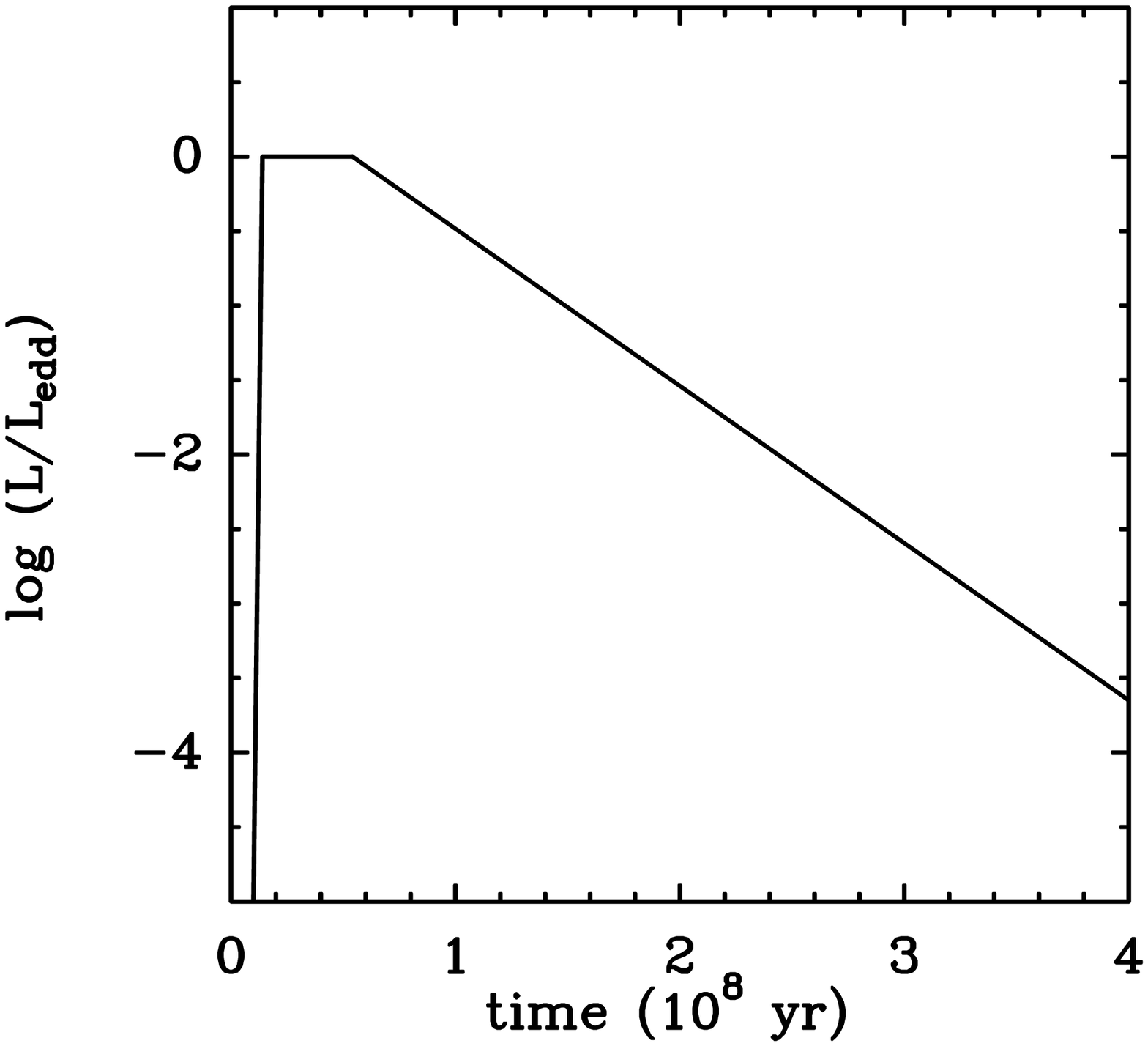}}\hfil
\epsfxsize=.45\textwidth \rotatebox{-90}{\epsfbox{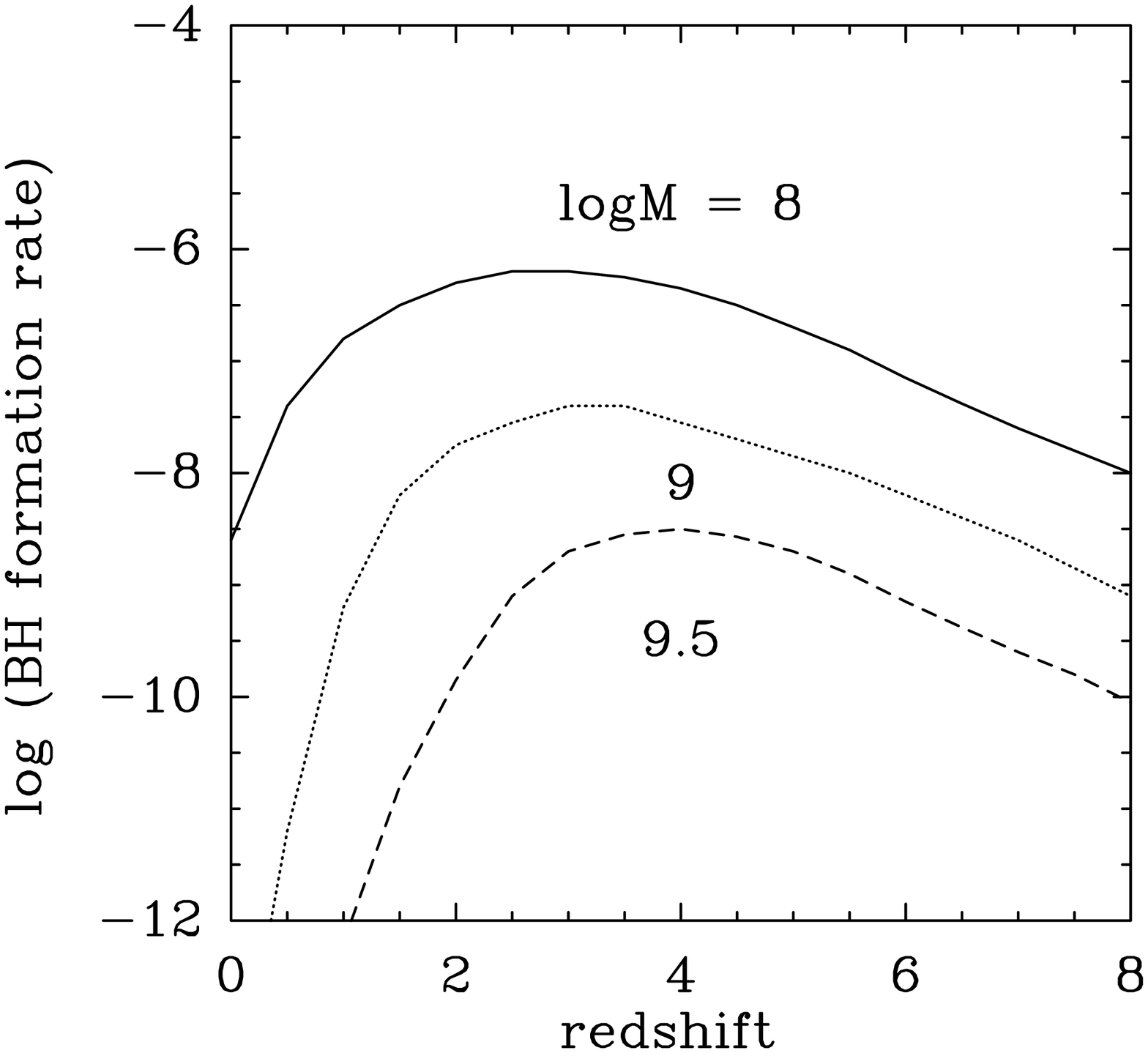}}}
\caption{  {\bf left} A schematic view of the activity phase of a typical
quasar: during the growth phase, the AGN does not radiate 
efficiently, although the fueling rate is larger than Eddington. 
After its luminous phase of  $\sim$ 4 10$^7$ yr, the fuel
is exhausted and the AGN fades away.
 {\bf right}  The black hole formation rate, or number density
of newly formed BH per comoving volume, per time, and lnM, 
in Mpc$^{-3}$ (10$^7$yr)$^{-1}$ (lnM)$^{-1}$
(from Haehnelt \& Rees 1993)}
\label{flash}
\end{figure}

The best interpretation of the observations lead to an AGN
activity life-time of $\sim$ 4 10$^7$ yr. During this
time the AGN would accrete and radiate at the Eddington limit.
Before, the black hole is not massive enough to accrete 
efficiently, since the Eddington limit is proportional to its mass. 
The external fueling rate is larger than the limit during the 
growth phase, then during its bright phase of $\sim$ 4 10$^7$ yr
the AGN radiates the most efficiently at Eddington luminosity,
until the fuel is exhausted, and the AGN fades away
(see fig \ref{flash}).

In CDM hierarchical scenarios, the smallest halos form first, and
they are denser, since the density of the Universe is a strong increasing 
function of redshift. Then the dynamical time-scale in such 
structures are much shorter. Due also to a higher gas mass fraction
at high redshift (not yet consumed in stars), there are many reasons
why the mass of the black holes formed are higher at high redshift
(even if they form in smaller-mass systems).
 This can explain the density distribution and luminosity function
observed for quasars (cf Boyle et al 1991): at high z, the quasars
were both more numerous and their black holes 
more massive (assuming that we
all see them in their bright phase, at Eddington luminosity). 

To take into account these arguments, and reproduce the fact that
the more massive black holes form in the smaller haloes, Haenelt
\& Rees (1993) assume the ratio:
$$
M_{bh}/M_{halo} \propto (1+z)^{6} exp(-v_*/vcir)^{3}
$$
the dependence in (1+z) corresponds to the higher efficiency
to form a black hole at high z (due to the high densities),
the exponential cut-off for low v$_{cir}$ takes
into account the difficulty of small mass system to retain gravitationnally
their gas in their small potential well ($v_*$ is a normalizing value).
With this model, the result is indeed that at low redshift, mostly small
black holes are built, while the peak of formation of very massive
black holes occurs at z=4 (see fig \ref{flash}).

Today, merging of galaxies create a quasar in only small-mass
gaseous objects. 
If the black hole formation process is not efficient at low redshift, 
the main process creating AGNs now is the tidally triggered
nuclear activity, through gas accretion and minor mergers.
To explain for instance the 1\% of Seyferts in field galaxies
(Huchra \& Burg 1992), it is sufficient to assume a duty cycle of
1\% of "normal" galaxies.

\begin{figure}[ht]
\begin{center}
\rotatebox{-90}{\includegraphics[width=.75\textwidth]{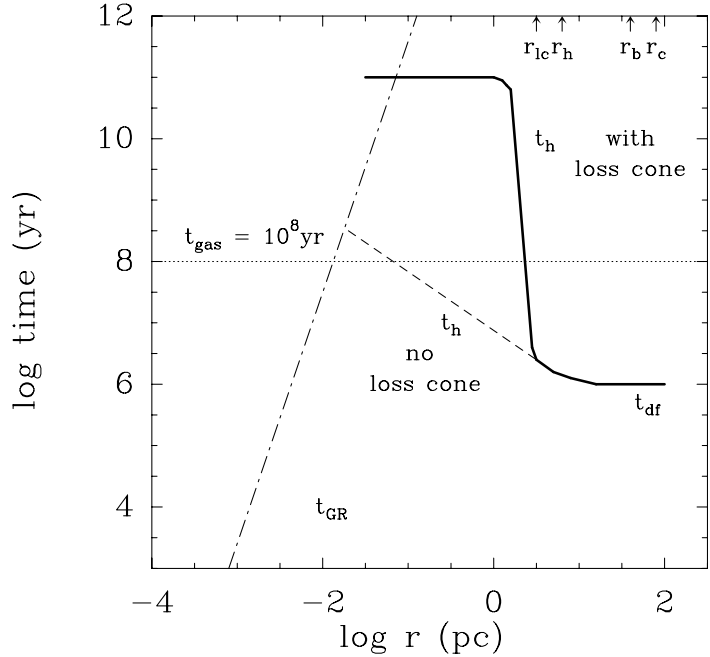}}
\end{center}
\caption{ Time-scales relevant to the formation of a black-hole
binary, in the merger remnant of two galaxies. Dynamical friction
in the core of the elliptical galaxy formed, is acting in $t_{df} \sim$
10$^6$ yr. Within $r_h$, the evolution will proceed on
$t_h = (r_h/r) t_{df}$ if loss-cone effects are neglected.
However, taking into account depletion of stars in the
loss cone leads to $t_h$ = 10$^{11}$ yr, a much longer time.
If infall of gas is considered, $t_{gas} \sim$ 10$^8$ yr
applies. At the end, gravitational radiation will
take over ($t_{GR}$); from Begelman et al (1980).
The core radius $r_c$, the ``bound'' radius $r_b$,
the hardening radius $r_h$ and the loss-cone radius
$r_{lc}$ are defined in the text.}
\label{begel}
\end{figure}

\subsection{Binary black hole formation}

Given the large frequency of galaxy encounters 
and mergers, if there is a massive black hole in nearly
every galaxy, the formation of a binary
black hole should be a common phenomenon.
The successive physical processes able to brake
the two black holes in their relative orbit have been
considered by Begelman et al (1980).
Each black hole sinks first toward the merger remnant center 
through dynamical friction onto stars. A binary is formed;
but the life-time of such a binary can be much larger
than a Hubble time, if there is not enough stars to
replenish the loss cone, where stars are able to interact
with the binary.
Once a loss cone is created, it is replenished 
only through the 2-body relaxation between stars,
and this can be very long (see section 2). 
Modelising the merger remnant as an elliptical,
with a core of radius $r_c$ and mass $M_c$
(and corresponding velocity $V_c$), the radius
where loss cone effects are significant is:
$r_{lc}/r_c=(M_{bh}/M_c)^{3/4}$.
The various time-scales involved, and 
corresponding characteristic scales are defined
by the following steps:
\begin{itemize}
\item the dynamical friction on stars, in less than a 
galactic dynamical time,
$$
t_{df} \sim  (V_c/300{\rm km/s}) (r_c/100{\rm pc})^2 (10^8/M_{bh}) \, {\rm Myr}
$$
\item when the separation of the binary shrinks to 
a value $r_b = r_c (M_{bh}/M_c)^{1/3}$, the two black holes become 
bound together
\item the binary hardens, with $r_h \propto (r/r_b)^{3/2}$
\item  but when $r < r_{lc} = (M_{bh}/M_c)^{3/4} r_c$,
 the stars available for the binary to interact with, are depleted through 
the loss cone effect, and replenished only by 
2-body relaxation
\item gas infall can reduce the binary life-time
(whether the gas is flung out, or accreted, there is 
a contraction of the binary) in $t_{gas}$
\item gravitational radiation shrinks the orbit on 
$t_{GR} \sim 0.3 {\rm Myr} (10^8/M_{bh})^3 (r/0.003{\rm pc})^4$
if the two black holes have comparable masses
\end{itemize}
All these time-scales are represented on figure \ref{begel}.
If the binary life-time is too long, another merger with
another galaxy will bring a third black-hole. Since a three-body
system is unstable, one of the three black-holes will be ejected
by the gravitational slingshot effect.

Since the life-time of the binary is not short,
there should be observable
manifestations of massive black hole binaries.
One of the best tracer is to detect the periodicity
of the keplerian motion, with
the period  P$\sim$ 1.6yr r$_{16}^{3/2}$ M$_8^{-1/2}$.
This is the case for the AGN
OJ 287 where eclipses have been monitored for a century 
(Takalo 1994, Lehto \& Valtonen 1996, Pietil\"a 1998).
Also, if the black holes are rotating, and their
spins have misaligned axes, they precess around 
the orbital one.
Plasma beams (aligned to the hole axis) precess,
and curved jets should be observed, 
with periods between 10$^3$ to 10$^7$ yr.
This is frequently the case in
radio structures observed with VLA and VLBI,
modified by Doppler boosting, and light travel time
(cf 3C 273, NGC 6251, 1928+738, Kaastra \& Roos 1992; 
Roos et al 1993).
Finally, pairs of radio galaxies have been observed during
their merger with four radio jets (3C75, Owen et al 1985).

Numerical simulations have brought more precision in
the determination of the life-time of the binary,
although numerical artifacts have given rise to debates.
Ebisuzaki et al (1991) claimed that the life-time of 
the binary should be much shorter if its orbit is excentric,
since then the binary can interact with more stars and
release the loss cone problem. The first numerical
simulations tended to show that orbit
excentricity should grow quickly through 
dynamical friction (Fukushige et al 1992).
Mikkola \& Valtonen (1992) and others found
that the excentricity in fact grows only very slowly.

Numerical simulations suffer from a restricted number
of bodies N, and consequently of a large random velocity of the
binary (that shoud decrease in N$^{-1/2}$).
The binary then  wanders in or even out of the loss cone,
and the effect of the loss cone depletion does not occur
(Makino et al 1993).
Also the 2-body relaxation time is shorter than
in the real system, contributing to replenish the cone.
Numerically, the life-time of the binary depends on the total 
number of particles, i.e. the ratio between the 
black hole to particle mass:
$$
t_b \propto (M_{bh}/m*)^{0.3}.
$$

To summarize the conclusions of several numerical
computations, there is finally
 little dependence on excentricity $e$, only in rare 
cases, when $e$ is large from the beginning 
(Quinlan 1996). 
Eventually, the wandering of the binary helps the merging
of the two black holes (Quinlan \& Hernquist 1997).
The ejection out of the core of stars interacting with the binary
weakens the stellar cusp, while the 
binary hardens. This may help to explain the
surprisingly weak stellar cusps in the center of giant ellipticals 
observed recently with HST.
Observations show that bright elliptical
galaxies have weak cusps, while faint 
galaxies have strong cusps, with a power law slope
of density versus radius of up to 2.
A way to weaken the cusps is a sinking black hole
(Nakano \& Makino 1999), and this could be the
case for giant galaxies that have experienced
many mergers in their life.

\begin{figure}
\begin{center}
\rotatebox{0}{\includegraphics[width=.75\textwidth]{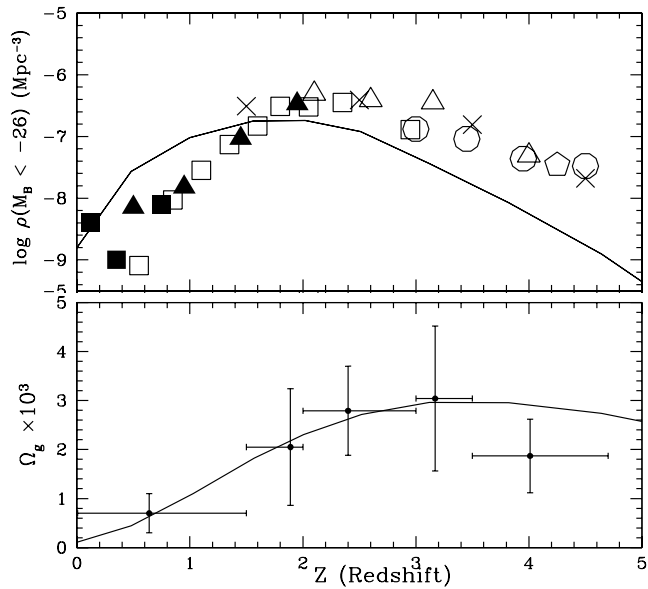}}
\end{center}
\caption{ 
 {\bf top}: The evolution of the space density of quasars 
with $M_B < -26$ for  the $\Lambda$CDM model
(from Kauffmann \& Haehnelt 2000).
{\bf bottom}: The cosmological mass density in cold gas 
in galaxies as a function of redshift for the $\Lambda$CDM model 
 (from Kauffmann \& Haehnelt 2000).
The data is taken from Storrie-Lombardi et al. (1996).}
\label{dla}
\end{figure}

\subsection{Hierarchical merging scenario}

Semi-analytic models, based on the Press-Schechter formalism, and a CDM 
hierarchical scenario of galaxy formation (Kauffmann \& Haehnelt 2000), can 
reproduce rather well the essential observations: the proportionality relation between
the bulge and black hole mass in every galaxy, the amount of energy radiated
over the Hubble time due to accretion onto massive black holes, the past
evolution of AGN activity. The assumptions are that the black holes 
grow through galaxy merging, both because of the merger of the 
pre-existing black holes, and due to the infall of gas to the center in
the merging, that can fuel the merged BH. It is also assumed that
the cold gas in galaxies decrease with time; this implies that the
fueling will also decrease with time, accounting for the observed decline of AGN
activity. Finally, the gas accretion time-scale is proportional
to the dynamical time-scale, which is shorter at high redshift. The quasars
convert mass to energy at a fixed efficiency, and cannot radiate
more than the Eddington limit.

The results of such simulations are a strong decrease of the gas fraction in
galaxies, from 75\% at z=3 to 10\% at z=0, corresponding to the gas density
decrease observed in the damped Lyman alpha systems (cf fig \ref{dla});
 this implies that the
black holes in the smallest ellipticals that formed at high z are relatively
more massive, since there was more gas at this epoch. Ellipticals forming
today have smaller black holes. Also the brightness of AGN for a given
galaxy was relatively higher in the past. The rapid decline of quasars
is then due to several causes: 

-- a decrease in the merging rate (which is also the cause of the 
decrease of the star forming rate)

-- the decrease of the gas amount in galaxies

-- the decrease of the accretion rate (the dynamical processes are  slower)

\noindent In these kind of models, it is natural to expect a ratio of proportionality
between bulge and black hole masses, since they both form from the same
mechanisms, the hierarchical merging, and the corresponding 
dynamical gas concentration. It is interesting to note that the life-time
of the quasar phase, a few 10$^7$ yr is then derived.

\section{Conclusions}

The source of fueling depends on the strength and 
luminosity of the AGN:

-- for low luminosity AGN and Seyferts, only stars from a dense 
nuclear cluster are sufficient (through tidal distorsions 
and stellar collisions),

-- for  high luminosity AGN and quasars, large accretion rates
are required, which involve large-scale gravitational
instabilities. Those drive gas towards the center
that trigger big starbursts,
and the coeval compact cluster just formed can provide 
the fuel through mass loss of young stars and supernovae.
This gas must have been driven from the 
galactic disks, through internal gravitational 
instabilities (bars, spirals), more generally the 
consequence of interactions and mergers.

Galaxy disks are in general far from stationary 
equilibrium, most often subject to $m=2$ and $m=1$ 
instabilities. These non-axisymmetric instabilities produce
gravity torques, that drive the gas inwards. When the
mass concentration towards the center is large enough,
a secondary bar can decouple and rotate with
a higher pattern speed. Embedded bars or
non-axisymmetric structures can take over the gas flow
to fuel the nucleus. The fueling is therefore 
favored in early-type objects. 

Interactions and mergers also produce bars and 
non-axisymmetric structures, that fuel the nucleus through their
gravity torques. They first trigger huge starbursts in the 
merger center, that could afterwards fuel the AGN. 
Since internal instabilities, external trigger and
hierachical merging both produce the bulges and fuel
the nucleus, it is natural to expect a proportionality
ratio in their masses.

\section*{Acknowledgments}
I am very grateful to the organisers for inviting me to give these 
lectures, in a very friendly and scientifically exciting ambiance.

\end{document}